\documentclass[twocolumn,apsfloatfix]{revtex4}
\usepackage[pdftex]{graphicx,color}
\usepackage{amsmath,amssymb,wasysym,graphicx,capt-of,ifthen,calc}
\usepackage{enumerate}
\usepackage{url}
\usepackage{subfigure}
\usepackage{float}
\begin{document}

\title{Reality Inspired Voter Models: A Mini-Review}
\author{S.~Redner}
\affiliation{Santa Fe Institute, 1399 Hyde Park Road, Santa Fe, NM, 87501
  USA}
\begin{abstract}
  This mini-review presents extensions of the voter model that incorporate
  various plausible features of real decision-making processes by
  individuals.  Although these generalizations are not calibrated by
  empirical data, the resulting dynamics are suggestive of realistic
  collective social behaviors.
\end{abstract}
\maketitle

\section{Introduction}\label{sec:intro}
The voter model (VM) is an idealized description for the evolution of
opinions in a population.  Each individual (voter) can assume one of two
states (e.g., $\mathbf{1}$/$\mathbf{0}$, $\uparrow/\downarrow$,
Democrat/Republican), and a single voter resides at each node of an arbitrary
network.  A voter is selected at random and adopts the state of a randomly
chosen neighboring voter.  Each individual is influenced only by a fixed set
of neighbors; there is no notion of a ``right'' or a ``wrong'' opinion, there
are no external influences (e.g., news media), or other types of
interactions.  The basic update step is repeated at unit rate until a
population of $N$ agents necessarily reaches consensus; this inevitability of
consensus is one of the basic features of the VM.

The simplicity and utility of the VM has sparked much research in probability
theory~\cite{CS73,HL75,C89,L99} and statistical
physics~\cite{K92,FK96,DCCH01,CFL09,KRB10,B18} to understand its dynamical
behavior.  One of the appeals of the VM is that it is exactly soluble when
voters are situated on regular lattices or homogeneous graphs.  However, it
is hard to envision that a large group of individuals can come to consensus
on anything.  This dichotomy between eventual consensus in the VM and the
common experience of opinion diversity has motivated generalizations of the
VM to include realistic aspects of opinion formation that can forestall
consensus.  Examples include individual stubbornness, partisanship, truth
versus falsehood, individual heterogeneity, social heterogeneity, and
multiple opinion states.

Some of these extensions of the VM are reviewed in this article; a recent
review that has some overlap with the topics discussed here is
Ref.~\cite{JS19}.  Unless stated otherwise, the underlying social network is
a complete graph of $N$ nodes.  While obviously unrealistic, this
geometry-free setting gives rise to much interesting phenomenology.
Additionally, real social networks are long ranged~\cite{FFF99,BKM00,N01}, so
the complete graph may be a better approximation than a regular lattice of
the underlying social substrate.  What is not treated here is the possibility
of changes in that social substrate at the same time that individual opinions
are changing.  For a discussion of such adaptive voter models see, e.g.,
Refs.~\cite{GDB06,HN06,KB08,SS08,SS10,DGL12,RG13}.

The focus of this article is primarily on topics that have been addressed by
my collaborators and myself; space limitations prevent a fuller discussion of
many important developments in this field.  While the social attributes
mentioned above are inspired by reality, the resulting models should still
not be viewed as realistic.  What is missing is a calibration of the
parameters of these more general models with real social data~\cite{GS18}.
In the absence of such a connection, these extensions of the VM might be
viewed as instructive theoretical games.  Nevertheless, they are
phenomenologically rich and perhaps hint at how to construct realistic
opinion dynamics models.

\section{Classic Voter Model}
\label{sec:classic}

\begin{figure}[ht]
\centerline{  \includegraphics[width=0.45\textwidth]{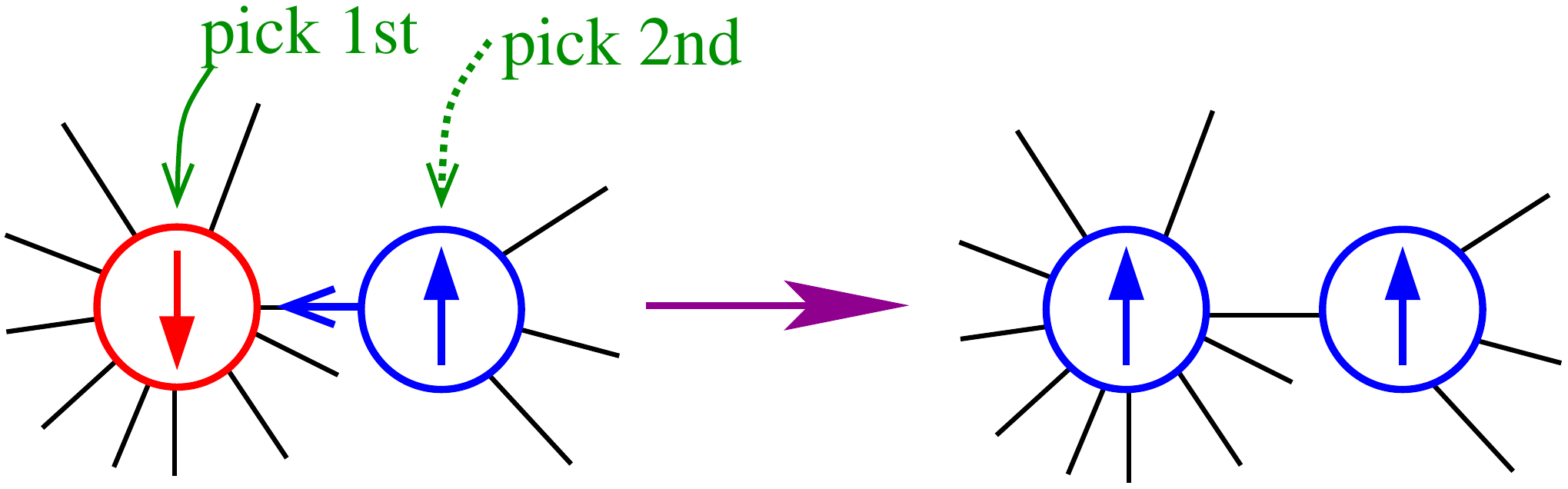}}
\caption{\small Voter model update rule on a graph: a random voter is picked
  and adopts the state of a random neighbor. }
\label{model}
\end{figure}

In the classic VM, $N$ voters live at the nodes of an arbitrary static graph,
with one voter per node.  The opinion evolution is simplicity itself
(Fig.~\ref{model}):
\begin{enumerate}
\itemsep -1ex
\item Pick a voter uniformly at random.
\item This voter adopts the state of a random neighbor.
\item Repeat 1 \& 2 until consensus is necessarily reached.
\end{enumerate}
Figuratively, each agent has no self-confidence and merely adopts the state
of one of its neighbors.  Two basic observables of the VM are: (i) the
\emph{exit probability} $E(m)$ that $N$ voters reach $\uparrow$ consensus as
a function of the initial magnetization
$m\equiv (N_\uparrow-N_\downarrow)/N$, where $N_\uparrow$ and $N_\downarrow$
are the initial number of $\uparrow$ and $\downarrow$ voters, respectively,
and (ii) the \emph{consensus time} $T(m)$, the average time to reach
consensus as a function of $m$.  For notational simplicity, the $N$
dependence is not written for $E(m)$ and $T(m)$

To mathematically express the update rule on a homogeneous graph, let
$\sigma_i=\pm 1$ denote the voter state at site $i$.  It is easy to check
that the rate $w_i$ at which this voter state $\sigma_i$ changes is given by
\begin{align}
  \label{w}
  w_i = \frac{1}{2}\Big(1-\frac{\sigma_i}{z}\sum_j \sigma_j\Big)\,,
\end{align}
where $z$ is the coordination number of each node and the sum runs over the
$j$ nearest neighbors of site $i$.  A crucial feature of this transition rate
is its linearity in the number of disagreeing neighbors.  This linearity
underlies the solvability of the voter model.  Another important consequence
of linearity is that there is no surface tension along an interface that
separates domains of $\uparrow$ and $\downarrow$ voters on finite-dimensional
lattices.  Here, surface tension means that the interaction tends to make
same-opinion clusters compact and therefore hard to break.  Thus the voter
model has a different character than the kinetic Ising model with single
spin-flip dynamics, in which interface motion is driven by surface
tension~\cite{DCCH01,KRB10,GSS83,B94}.

Using \eqref{w} is immediate to verify that the average magnetization,
$m\equiv \sum_i \langle\sigma_i\rangle/N$, is conserved.  The time evolution
of the average of $\sigma_i$ is given by
\begin{subequations}
  \label{dot-sigma}
\begin{align}
  \langle \dot \sigma_i\rangle &= -2\langle \sigma_i w_i\rangle\,,
\end{align}
because whenever a voter flips, with rate $w_i$, the change in the value of
the voter's state is $-2\sigma_i$.  Substituting the flip rate \eqref{w} into
the above equation gives
\begin{align}
  \label{dot-sigma-b}
  \langle \dot \sigma_i\rangle & = -\langle \sigma_i\rangle
      + \frac{1}{z}\sum_j \langle \sigma_j\rangle\,,
\end{align}
\end{subequations}
where we use the fact that $\sigma_i^2=1$.  Summing \eqref{dot-sigma-b} over
all sites $i$ gives magnetization conservation, $\dot m=0$.

This conservation law immediately determines the exit probability.  When the
initial magnetization equals $m$, the average value of the final
magnetization when consensus is eventually reached equals
\begin{align}
  \label{m-inf-VM}
  \langle m_\infty\rangle = 1\times E(m)+ (-1)\times \big(1-E(m)\big)\,.
\end{align}
That is, a final magnetization $+1$ is reached with probability $E(m)$, while
$m_\infty=-1$ is reached with probability $1-E(m)$.  Since
$\langle m_\infty\rangle =m$ by magnetization conservation, we have
$E(m)=\frac{1}{2}(1+m)$.

There is a systematic tool with many applications to first-passage
processes~\cite{K31,K97,R01}---the backward Kolmogorov equation---that can be
used to determine both the exit probability and the consensus time in a
simple way.  We will use this formalism throughout this article.  Suppose
that the density of $\uparrow$ voters equals $\rho$.  After a single update,
$\rho$ may change to $\rho\pm \delta\rho$ or remain the same, with respective
probabilities $w_{\rho\to \rho\pm \delta\rho}$ and $w_{\rho\to \rho}$, after
which the process restarts.  Accordingly, the exit probability obeys the
backward Kolmogorov equation
\begin{subequations}
  \label{ET-def}
\begin{align}
  \label{E-def}
  E(\rho) &= w_{\rho\to \rho+\delta\rho} E(\rho+\delta\rho)\nonumber\\
  &+   w_{\rho\to \rho-\delta\rho} E(\rho-\delta\rho)
  +    w_{\rho\to \rho}E(\rho)\,,
\end{align}
subject to the boundary conditions $E(0)=0$, $E(1)=1$.  For the VM, the
flipping probabilities are $w_{\rho\to \rho\pm\delta\rho}=\rho(1-\rho)$ and
$w_{\rho\to \rho}= 1 -2\rho(1-\rho)$.  Taking the continuum limit gives
$d^2E/d\rho^2=0$, with solution $E(\rho)=\rho$, or equivalently,
$E(m)=\frac{1}{2}(1+m)$.  It is purely a matter of convenience whether to use
$m$ or $\rho$ as the dependent variable.

We obtain the time to reach either $\uparrow$ or $\downarrow$ consensus by
the direct analog of Eq.~\eqref{E-def}:
\begin{align}
\label{T-def}
  T(\rho) &= w_{\rho\to \rho+\delta\rho}\big[ T(\rho\! +\!\delta\rho) +\delta t\big]\nonumber\\
      &     + w_{\rho\to \rho-\delta\rho}\big[ T(\rho\!-\!\delta\rho)+\delta t\big]
           + w_{\rho\to \rho}\big[T(\rho)+\delta t\big]\,.
\end{align}
\end{subequations}
That is, the consensus time from an initial state with density of $\uparrow$
voters equal to $\rho$ is the probability for an update to a new state where
one voter has changed its opinion, multiplied by the time to reach consensus
from this new state plus the time $\delta t=1/N$ for a single update.  The
continuum limit now gives $N\rho(1-\rho)d^2\,\, T/d\rho^2 = -1$, subject to
the boundary conditions $T(0)\!=\! T(1)\!=\! 0$.  The solution is
\begin{equation}
\label{VM:KG:contime:sol}
T(\rho)= -N\big[(1-\rho)\ln(1-\rho)+\rho\ln \rho\big] \,.
\end{equation}
This linear $N$ dependence also represents the generic behavior for the
consensus time of the VM on Euclidean lattices in spatial dimensions
$d\geq 3$.  For $d=2$, $T\sim N\ln N$, while for $d=1$,
$T\sim N^2$~\cite{CS73,HL75,C89,L99,K92}.

\section{Stubborn/Confident Voters}

\subsection{Individual Heterogeneity}

While each individual in the VM has no self confidence, real people do have
some of this attribute, and it is natural to examine its role on VM dynamics.
Self confidence can be viewed as quantifying the relative weights of
individual versus social information.  Other factors that influence the
weights of these two information sources include: lower cost of individual
versus social information, prestige or authority of an acquaintance,
propensity for conformity, perceived differences in accuracy of individual
versus social information, etc.  Many of these issues have been explored in
the social science literature (see, e.g., \cite{A51,KBR18}).

Perhaps the simplest way to implement self confidence is to endow each voter
$i$ with its own intrinsic flip rate $r_i$~\cite{MGR10}.  That is, when a
voter consults a neighbor with a different opinion, the voter changes state
with rate $r_i$ instead of with rate 1.  Such a diversity of individual
thresholds to a stimulus has been found to play an important role in
controlling collective social behaviors~\cite{G78,W02,M08}.  The transition
rate of voter $i$ with intrinsic flip rate $r_i$ in this \emph{heterogeneous}
voter model (HVM) now is (compare with Eq.~\eqref{w})
\begin{align}
  \label{w-het}
  w_i = \frac{1}{2}\Big(1-\frac{r_i\sigma_i}{z}\sum_j \sigma_j\Big)\,,
\end{align}
Following the same steps as in Eq.~\eqref{dot-sigma}, we may verify that it
is not the magnetization, but rather the inverse rate-weighted magnetization,
$\omega \equiv\langle \sigma_i/r_i\rangle$, which is conserved.  As a
consequence, the probability for a population with initial value $\omega$ to
reach $\uparrow$ consensus equals $\omega$.  Thus a small fraction of
$\uparrow$ voters with very small flip rates---stubborn voters---leads to a
probability of reaching $\uparrow$ consensus that can be arbitrarily close to
1, even if most of the population is initially in the $\downarrow$ state.

One can infer the average consensus time $T$ for $N$ heterogeneous voters on
the complete graph by the following simple argument.  For specificity,
suppose that the flip rate distribution is $p(r)=A\,r^{-\alpha}$, with
$r\in (0,r_{\rm max}]$ and $\alpha\in [0,1)$ so that $p(r)$ is normalizable.
To compare across different $\alpha$ values, we fix the average flip rate of
the population to be 1.  These conditions give
$r_{\rm max}= {(2-\alpha)}/{(1-\alpha)}$ and
$A=(2-\alpha) r_{\rm max}^{\alpha-2}$.

A voter with flip rate $r$ needs to attempt of the order of $\frac{1}{r}$
flips before it actually changes state.  Consequently, we can view the
population as containing an effective number of voters $N/r$.  Since the
consensus time of the VM is proportional to $N$, we anticipate that the
consensus time in the HVM should be proportional to the average value of
$N/r$.  Also, as will be shown in Eq.~\eqref{TRHS}, the consensus time of the
voter model on a general complex graph is proportional to the effective
population size.  With this hypothesis, the consensus time in the HVM is
$T\sim N\langle 1/r\rangle$.  To determine $\langle 1/r\rangle$ for the
flip-rate distribution $p(r)=Ar^{-\alpha}$, we first use the fact that the
smallest rate $r_{\rm min}$ among a finite population of $N$ voters is
non-zero and determined by the extremal criterion~\cite{G87}
\begin{equation*}
  \int_0^{r_{\rm min}}\!\! A\,r^{-\alpha}\,dr=N^{-1}\,,
\end{equation*}
i.e., one voter in a population of $N$ has a flip rate $r_{\rm min}$ or
smaller.  This criterion gives $r_{\rm min}\sim N^{-1/(1-\alpha)}$.  Now we
estimate $\langle 1/r\rangle$ as
\begin{align*}
  \langle 1/r\rangle&=\int_0^{r_{\rm max}} \frac{p(r)}{r}\,dr \sim
                      \int_{r_{\rm  min}}^{r_{\rm max}} 
  \frac{p(r)}{r}\,dr \sim N^{\alpha/(1-\alpha)}\,.
\end{align*} 
Finally, 
\begin{equation}
\label{TN}
T \sim N^{1/(1-\alpha)}\,,
\end{equation}
for $0<\alpha<1$, while $T \sim N\ln N$ for $\alpha=0$.  Thus heterogeneity
in individual flip rates forestalls consensus, as the the consensus time
scales superlinearly with $N$.  Notice that the inverse rate of the
stubbornest voters is the same order as the consensus time, so it is these
stubbornest voters that control the approach to consensus.  The phenomenon
that a small fraction of stubborn individual can overcome the majority
opinion has been studied both in the social science~\cite{M80,M85} and
physics literatures~\cite{GJ07,XSK11}, and begins to address the basic
question of how can a small minority opinion group can eventually supplant a
majority opinion.

\subsection{Confident Voting}

The lack of individual voter confidence is one of many unrealistic aspects of
the VM.  There are a variety of socially/cognitively plausible mechanisms
that people might use to solve the dilemma of how much weight to give to self
knowledge compared to the knowledge of others when deciding what to believe
or what to do.  One such mechanism is reinforcement; in the context of the
voter model, a voter night typically require multiple encounters with
neighbors that hold a different opinion before the voter actually changes
opinion.  This feature was also highlighted in the social experiment of
Centola~\cite{C10}, where individual behavior changed only when a person
received multiple reinforcing inputs from others.  Such threshold behavior
also arises in the $q$-voter model~\cite{CMP09} and in contagion
models~\cite{DW04}.

\begin{figure}[ht]
\centerline{\includegraphics*[width=0.45\textwidth]{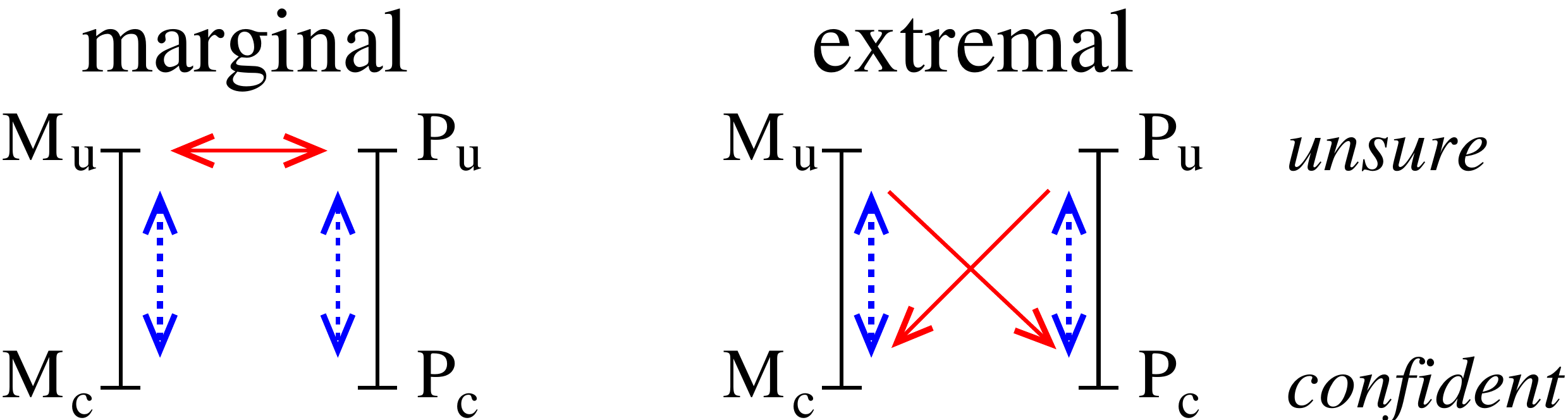}}
\caption{\small Illustration of the (a) marginal, and (b) extremal confident
  voter models.  Dashed arrows indicate possible confidence level changes,
  while solid arrows indicate possible opinion change events.}
  \label{transitions-conf}
\end{figure}

We investigate the role of a threshold through the \emph{confident} voter
model (CVM)~\cite{VR12}, in which each voter has two opinion states and two
levels of commitment to an opinion--—confident and unsure.  Upon interacting
with an agent of a different opinion, a confident voter becomes unsure but
keeps its opinion, while an unsure agent changes opinion.  That is, a
confident voter requires two consecutive prompts before changing opinion in
the CVM (Fig.~\ref{transitions-conf}).

The basic variables are the opinion of each voter and its confidence level.
We label the states of an agent as $P_c$ and $P_u$ for confident and unsure
$\uparrow$ agents, respectively, and correspondingly $M_c$ and $M_u$ for
$\downarrow$ agents.  There are two natural variants of confident voting:
\begin{itemize}
\itemsep -0.25ex

\item \emph{Marginal:} An unsure agent that changes opinion is also unsure
  about the new opinion state and can switch back in a single update.

\item \emph{Extremal:} An unsure agent ``sees the light'' after an opinion
  change and becomes confident of its opinion.  This voter again requires two
  interactions with an opposite-opinion voter to switch another time.

\end{itemize}

With the assumption that the rates of all processes sketched in
Fig.~\ref{transitions-conf} equal 1, it is simple exercise to write the rate
equations for the densities of voters in each state.  For the marginal
version, these equations are:\hfil\break\medskip

\emph{Marginal:}
\begin{subequations}
\begin{align}
\label{REM}
  \begin{split}
\dot P_c&= -(M_c+M_u)P_c+P_cP_u \,,\\
\dot P_u&=  ~~MP_c-P_cP_u+(M_uP_c-M_cP_u)\,,
\end{split}
\end{align}
with parallel equations for $M_c$ and $M_u$ by interchanging
$M\leftrightarrow P$.  For the extremal version, the rate equations are (and
similarly for $M_c$ and $M_u$)\hfil\break\medskip

\emph{Extremal:}
\begin{align}
  \label{REE}
  \begin{split}
\dot P_c&= -M_cP_c+ M_uP_u+P_cP_u\,,\\
\dot P_u&= ~~ M_cP_c-M_uP_u-P_cP_u+(M_uP_c\!-\!M_cP_u)\,.
\end{split}
\end{align}
\end{subequations}

\begin{figure}[ht]
  \centerline{\includegraphics*[width=0.25\textwidth]{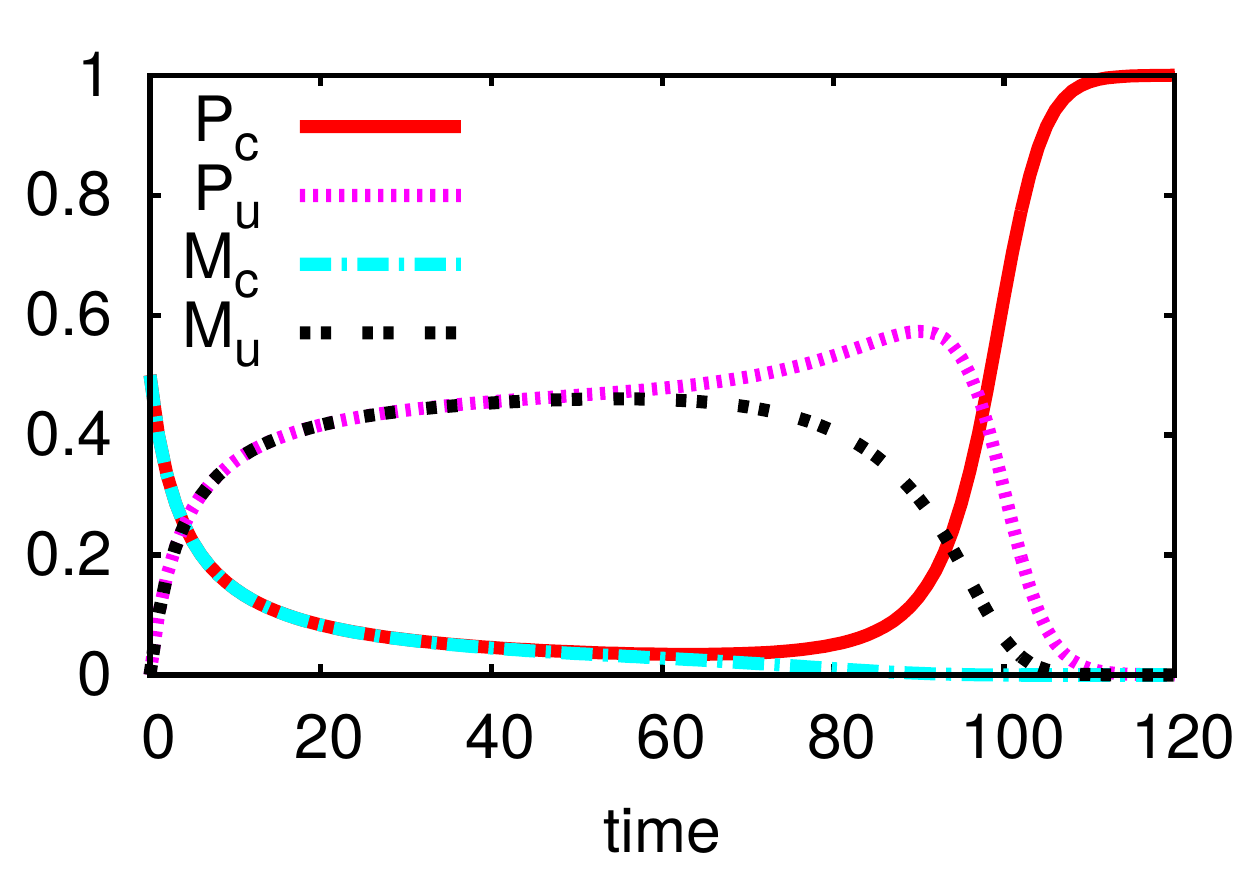}
    \includegraphics*[width=0.25\textwidth]{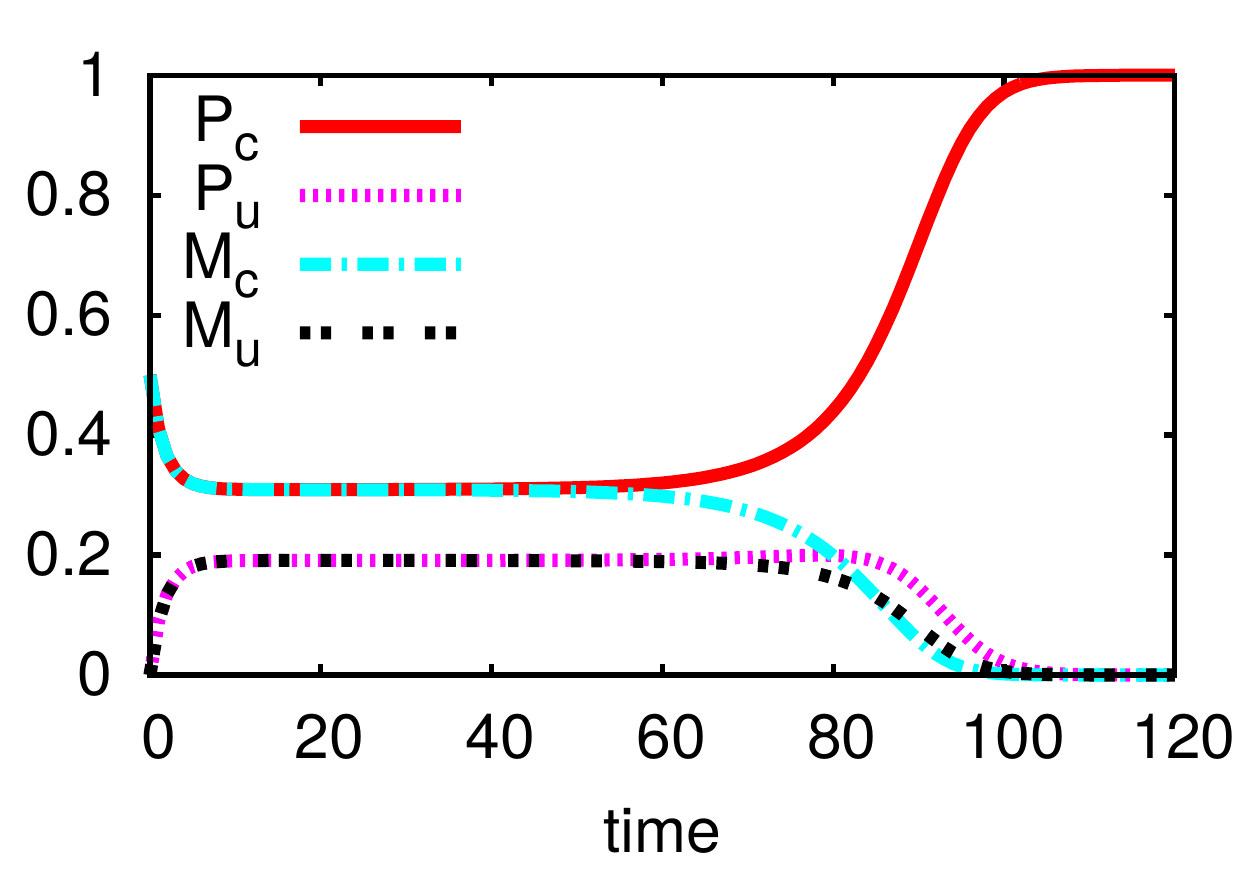}}
  \caption{\small Evolution of the densities for the marginal (left) and
    extremal (right) models with the near-symmetric initial condition
    $P_c=0.50001$, $M_c=0.49999$, and $P_u=M_u=0$.}
  \label{non-symm}
\end{figure}

Solving these equations~\cite{VR12} gives the following basic results: For a
symmetric population with equal densities of $\uparrow$ and $\downarrow$
voters, the final densities are $P_c=M_c=0$ and $P_u=M_u=1$ for the marginal
model and $P_c=M_c=\frac{1}{2}-P_u=\frac{1}{2}-M_u=\frac{1}{4}(\sqrt{5}+1)$
for the extremal model.  It is intriguing that for the extremal version of
the CVM, a symmetric population does \emph{not} reach consensus.  However,
for a slight asymmetry in the initial conditions, consensus is reached after
a time scale that is of the order of $\ln N$.  Intriguingly, the relaxation
to consensus is governed by two widely separated time scales
(Fig.~\ref{non-symm}).  We will encounter another example of this intriguing
multiple time-scale relaxation in a 3-state voter model with constrained
voting rules in Sec.~\ref{subsec:3-choice}.

\section{HETEROGENEOUS NETWORKS}
\label{sec:network}

A fruitful extension of the voter model is to networks with broad
distributions of degrees~\cite{SEM04,SEM05,CLB05,SR05,ARS06,SAR08,VE08}; the
degree of a node is the number of incident links on this node.  Here, the
magnetization is no longer conserved and the route to consensus is
dramatically different than for the VM on degree-regular networks.

\begin{figure}[ht]
  \vspace*{0.cm}
  \includegraphics*[width=0.3\textwidth]{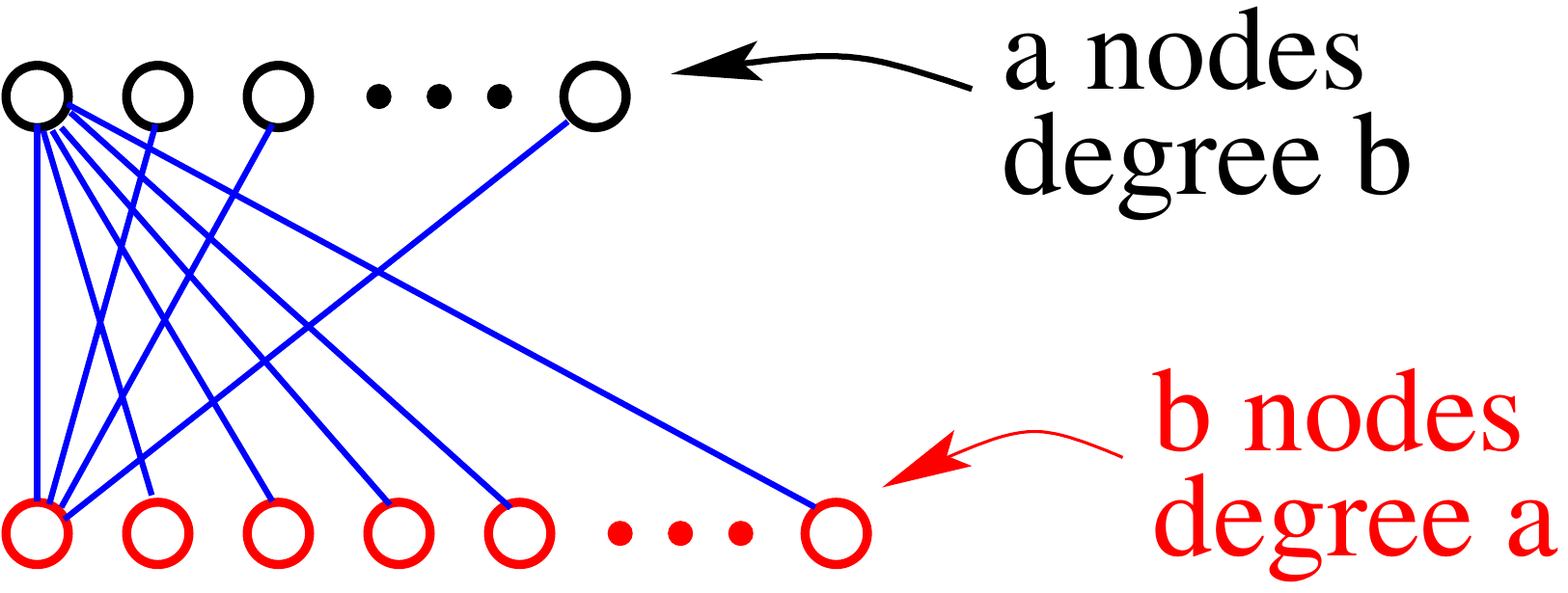}
 \caption{\small The complete bipartite graph $K_{a,b}$.   }
 \label{Kab}
\end{figure}

A useful way to begin to understand the role of degree heterogeneity on VM
dynamics is to study the simplest graph with different degrees, namely, the
complete bipartite graph $K_{a,b}$~\cite{SR05}.  This graph consists of two
subgroups, in which each member of subgroup $a$ is connected to every member
of subgroup $b$ and vice versa.  Thus $K_{a,b}$ consists of $a$ nodes of
degree $b$ and $b$ nodes of degree $a$ (Fig.~\ref{Kab}).  On this network, it
is a simple exercise to solve the VM dynamics.  Let $N_{a,b}$ be the
respective number of $\uparrow$ voters on each subgraph.  In an update event,
these numbers change according to
\begin{align}
\label{RE-N}
dN_a &= \frac{a}{a+b}\Big[\frac{a-N_a}{a}\frac{N_b}{b}-\frac{N_a}{a}\frac{b-N_b}{b}\Big]\,, \nonumber\\
dN_b &= \frac{b}{a+b}\Big[\frac{b-N_b}{b}\frac{N_a}{a}-\frac{N_b}{b}\frac{a-N_a}{a}\Big]\,.
\end{align} 
The gain term in $dN_a$ accounts for flipping a $\downarrow$ voter in
subgraph $a$ due to its interaction with a $\uparrow$ voter in $b$, while the
loss term accounts for flipping a $\uparrow$ voter in $a$.  Since the time
increment for an event $\delta t= 1/(a+b)=1/N$, the subgraph densities
$\rho_a=N_a/a$ and $\rho_b=N_b/b$ obey
$\dot\rho_{a,b} = \rho_{b,a}-\rho_{a,b}$, with solution
\begin{align}
\label{rhoav}
  \rho_{a,b}(t)=\tfrac{1}{2}[\rho_{a,b}(0)\!-\!\rho_{b,a}(0)]\, e^{-2t}
  + \tfrac{1}{2}[\rho_a(0)\!+\!\rho_b(0)].
\end{align}

Thus $\rho_a\!+\!\rho_b$ is asymptotically conserved and approaches
$\rho_\infty\equiv \frac{1}{2} [\rho_a(0)+\rho_b(0)]$, but the magnetization
$m=(a\rho_a+b\rho_b)/(a+b)$ is not conserved~\cite{SEM04}.  Because
$\rho_a+\rho_b$ is the conserved quantity, the sum of the subgraph densities
in the final state equals 2 with probability $E(\rho_a,\rho_b)$, which is the
exit probability to the state where all voters are in the $\uparrow$ state.
Conservation therefore gives
\begin{eqnarray}
\label{EP}
  E(\rho_a,\rho_b)  =  \tfrac{1}{2}[\rho_a(0) + \rho_b(0)]\,.
\end{eqnarray}
When the voters on the two subgraphs are initially oppositely oriented, there
is an equal probability of ending with all $\uparrow$ voters or all
$\downarrow$ voters, {\it independent\/} of the subgraph sizes.  In the
extreme case of the star graph $K_{a,1}$, with $a\gg 1$ voters with
$\uparrow$ opinion at the periphery and a single $\downarrow$ voter at the
center, there is a 50\% chance of $\downarrow$ consensus.

We follow the approach outlined in Sec.~\ref{sec:classic} to determine the
consensus time $T(\rho_a,\rho_b)$.  This time satisfies the backward
Kolmogorov equation~\cite{K31,K97,R01,SR05},
\begin{align}
\label{T-REC-BIP}
T(\rho_a,\rho_b)&\!=\!\nonumber
w(\rho_a,\rho_b\to\rho_a\!\pm\!\tfrac{1}{a},\rho_b)
[T(\rho_a\!\pm\!\tfrac{1}{a},\rho_b)+\delta t]\nonumber\\
&\!+\!w(\rho_a,\rho_b\to\rho_a,\rho_b\!\pm\!\tfrac{1}{b})
[T(\rho_a,\rho_b\!\pm\!\tfrac{1}{b})+\delta t]\nonumber\\
&\!+\!w(\rho_a,\rho_b\to\rho_a,\rho_b)
[T(\rho_a,\rho_b)+\delta t],
\end{align}
which generalizes Eq.~\eqref{T-def} to the complete bipartite graph.  The
first term (which is actually a shorthand for the two contributions with a
$+$ sign and a $-$ sign) accounts for flipping a $\downarrow$ ($\uparrow$)
voter in subgraph $a$ so that $\rho_a\to\rho_a\pm \frac{1}{a}$.  The
probability for flipping a $\downarrow$ voter in subgraph $a$ is
$w(\rho_a,\rho_b\to\rho_a+\frac{1}{a},\rho_b) =
\frac{a}{a+b}\,(1-\rho_a)\rho_b$, where $\frac{a}{a+b}(1-\rho_a)$ is the
probability to choose a $\downarrow$ voter in subgraph $a$.  Similar
explanations apply for the other terms in Eq.~\eqref{T-REC-BIP}.  This
equation is subject to the boundary conditions $T(0,0)=T(1,1)=0$.

\begin{figure}[ht]
\vspace*{0.cm}
\includegraphics*[width=0.425\textwidth]{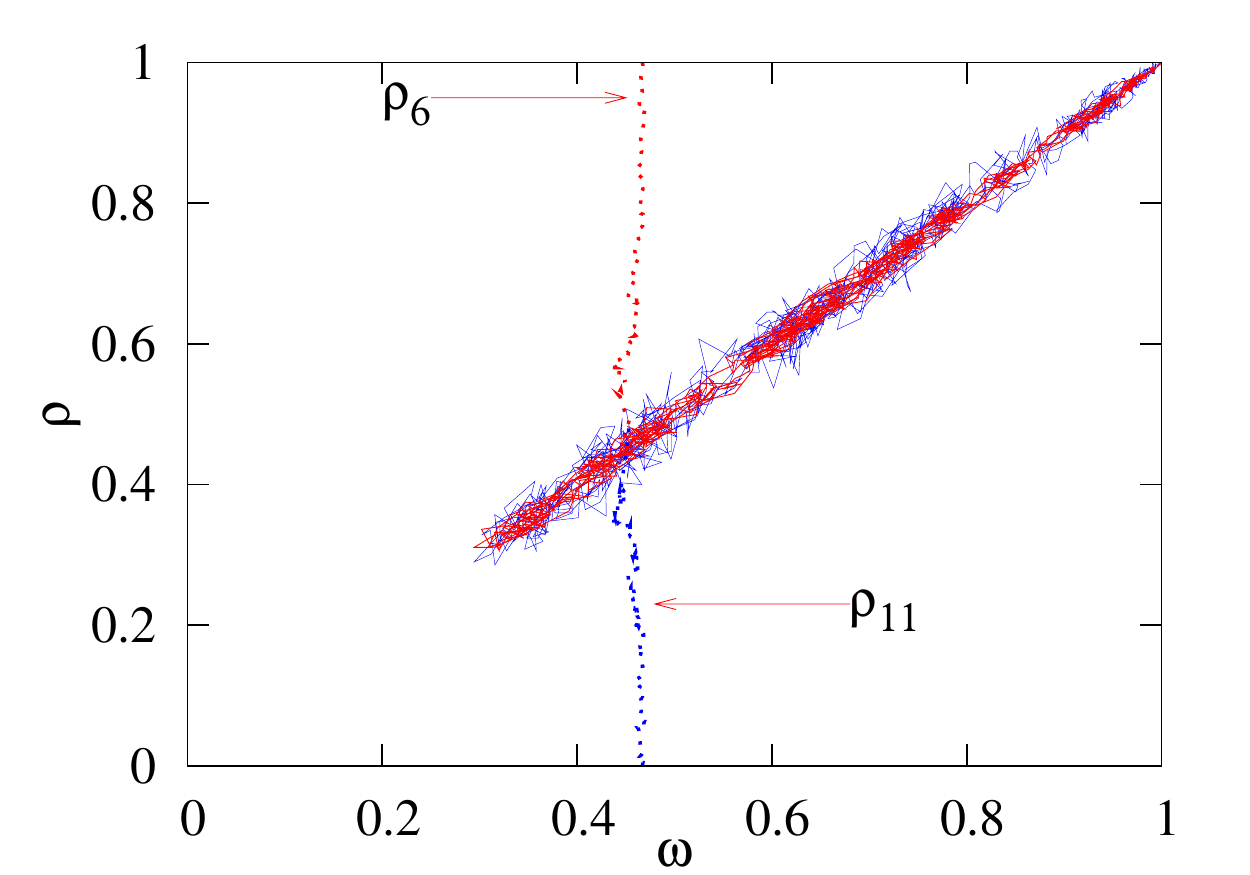}
\caption{\small Trajectories of $\rho_{6}(t)$ (degree less than $\mu_1=8$)
  and $\rho_{11}(t)$ (degree greater than $\mu_1$) versus $\omega$, for one
  realization of the voter model on a configuration model network of
  $2 \times 10^5$ nodes, with degree distribution $n_k\sim k^{-2.5}$, and
  average degree $\mu_1=8$.  The initial state is
  $(\rho_{k>\mu_1},\rho_{k\leq\mu_1}) =(0,1)$.  The dotted curves show the
  initial transient for $t\alt 1$, after which diffusive motion leads to
  consensus at $(1,1)$.}
\label{BGEvo}
\end{figure}

Expanding this recursion to second order, gives, after straightforward
algebra,
\begin{eqnarray}
\label{Bip-CT}
N\delta t &=& (\rho_a-\rho_b)(\partial_a-\partial_b)T(\rho_a,\rho_b) \\
&& - \tfrac{1}{2}(\rho_a+\rho_b-2\rho_a\rho_b)\left( \tfrac{1}{a}\,\partial_a^2 +
 \tfrac{1}{b}\,\partial_b^2\right)T(\rho_a,\rho_b)\nonumber
\end{eqnarray}
where $\partial_i$ denotes partial derivative with respect to $\rho_i$.  The
first term on the right corresponds to a convection that drives the
population to equal subgraph magnetizations in a time scale of order one,
while the second term corresponds to the subsequent diffusive fluctuations
that govern the ultimate approach to consensus (Fig.~\ref{BGEvo}).  We thus
determine the consensus time by replacing the subgraph densities $\rho_a$ and
$\rho_b$ by their common value $\rho$.  Doing so ignores early-time
transients when the subgraph densities are unequal.  We also transform the
derivatives with respect to $\rho_a$ and $\rho_b$ in Eq.~\eqref{Bip-CT} to
derivative with respect to $\rho$ to yield
\begin{subequations}
  \label{t-all}
\begin{equation}
\label{t-eqn}
\tfrac{1}{4}\,\rho(1-\rho)\left(\tfrac{1}{a}+\tfrac{1}{b}\right)\partial^2 T = -1\,,
\end{equation}
with solution
\begin{equation}
T_{N}(\rho) = -\frac{4 a b}{a+b}\left[ (1\!-\!\rho)\ln(1\!-\!\rho) + \rho\ln\rho\,\right]\,.
\end{equation}
\end{subequations}
Notice the close correspondence between the results given in
Eqs.~\eqref{t-all} with Eq.~\eqref{VM:KG:contime:sol} and its defining
differential equation.  For $a = {\cal O}(1)$ and $b = {\cal O}(N)$ (star
graph), the consensus time $T_{N}\sim{\cal O}(1)$, while if both $a$ and $b$
are ${\cal O}(N)$, then $T_{N}\sim {\cal O}(N)$, as on a complete graph.

This approach can be extended in a natural way to networks with an arbitrary
distribution of degrees.  If we neglect correlations between the degrees of
neighboring nodes, then Eq.~\eqref{T-REC-BIP} readily generalizes from two
densities on nodes of degrees $a$ and $b$ to an unlimited number of densities
$\rho_k$ on the set of nodes of any degree $k$.  Now taking the continuum
limit of this equation and making use of the conservation law that the
degree-weighted density of $\uparrow$ voters~\cite{SR05}
\begin{equation*}
  \omega \equiv \sum_k k n_k \rho_k/\sum_k k n_k
\end{equation*}
is conserved, we eventually arrive at the counterpart
of Eq.~\eqref{t-eqn}, namely,
\begin{equation}
\frac{1}{N\mu_1^2}\sum_k k^2 n_k\,\omega(1-\omega)\,\,\partial^2_\omega T =-1\,,
\end{equation}
with $\mu_m = \sum_k k^m n_k$ the $m^{\rm th}$ moment of the degree
distribution.  The solution for the consensus time is (compare again with
Eq.~\eqref{VM:KG:contime:sol})
\begin{align}
\label{TRHS}
T(\omega) &= -N \frac{\mu_1^2}{\mu_2}\,\,[(1-\omega)\ln(1-\omega)+\omega\ln\omega\,]\nonumber\\
&\equiv N_{\rm eff} [(1-\omega)\ln(1-\omega)+\omega\ln\omega\,,
\end{align}
With $N_{\rm eff}$ the \emph{effective} population size.

For a scale-free network whose degree distribution is given by
$n_k\sim k^{-\nu}$, the $m^{\rm th}$ moment of this distribution is given by
\begin{align*}
  \mu_m\sim \int ^{k_{\rm max}} k^m n_k\, dk\,.
\end{align*}
Here $k_{\rm max}\sim N^{1/(\nu-1)}$ is the maximal degree in a finite
network of $N$ nodes, which obtained from the extremal condition
$\int_{k_{\rm max}} k^{-\nu}\, dk = N^{-1}$ \cite{G87}.  Notice that the
second moment diverges at the upper limit for $\nu\leq 3$ while the first
moment diverges for $\nu\leq 2$.  Assembling these results for the moments,
the mean consensus time on a scale-free graph has the $N$ dependence
\begin{equation}
\label{Tf}
T_N\sim
\begin{cases}
N  & \qquad \nu>3,\cr
N/\ln N & \qquad  \nu=3,\cr
N^{(2\nu-4)/(\nu-1)} & \qquad  2<\nu<3,\cr
(\ln N)^2 &  \qquad \nu=2,\cr
{\cal O}(1) & \qquad  \nu<2.
\end{cases}
\end{equation}
The primary result is that consensus is achieved quickly on networks with
broad degree distributions.  This rapid consensus is facilitated by
``hubs''---nodes of the largest degrees that influence their very many
neighbors.  Moreover, there is a two time-scale route to consensus, with a
rapid approach to a state where the densities of $\uparrow$ voters on nodes
of any degree reach a common value (Fig.~\ref{BGEvo}), after which diffusive
fluctuations drive the approach to consensus.

\section{Non-Conserved Dynamics}

\subsection{Majority Rule} 
\label{subsec:maj}

While the VM update rule is simple and natural, there are other plausible
ways in which a group of voters might change opinions.  A basic example is
\emph{majority rule}, for which there exists a vast social science literature
(see e.g., \cite{C85,GOF83,BR05,CL09}).  From the mathematical perspective,
there are two natural implementations of majority rule.  One is that a voter
adopts the state of the majority in its local neighborhood.  This is just the
update rule of the kinetic Ising model with single spin-flip dynamics at zero
temperature~\cite{GSS83,B94}.  This well-studied dynamics leads to a
coarsening mosaic of $\uparrow$ and $\downarrow$ domains that usually, but
not always (see e.g., \cite{SKR2001a,SKR2001b}), ends in consensus.

Alternatively, a group of voters is selected and \emph{all} voters in this
group adopt the local majority opinion.  Related models have been studied in
various contexts~\cite{G99,SS00,G02,D02}.  Here we treat the situation where
each voter has two opinion states and the group size equals three---the
smallest possible size of a group with an odd number of neighbors.  In an
update, a group of three voters is selected at random from a total population
of $N$, corresponding to the complete-graph or mean-field limit.  These
voters all adopt the group majority opinion and return to the general
population~\cite{KR03}.  This update is repeated until a finite population
necessarily reaches consensus.

Let us first determine $E_n$, the exit probability that the population ends
with all $\uparrow$ voters when starting with $n$ $\uparrow$ voters.  To
compute $E_n$, note first that the quantity
\begin{eqnarray*}
\binom{3}{j}\binom{N-3}{n-j}\Big/\binom{N}{n}
\end{eqnarray*}
is probability that a randomly chosen group of three voters has $j$ voters of
$\uparrow$ opinion and $3\!-\!j$ voters of $\downarrow$ opinion when an
$N$-voter population contains $n$ $\uparrow$ voters.  This group becomes all
$\uparrow$ for $j=2$, becomes all $\downarrow$ for $j=1$, while for $j=0$ or
3 there is no evolution.  Thus $E_n$ obeys the backward Kolmogorov equation
\begin{subequations}
\begin{align}
\label{ME}
E_n&=\left\{ 3\binom{N\!-\!3}{n\!-\!2} E_{n+1}
\!+\!3\binom{N\!-\!3}{n\!-\!1}E_{n-1}\right.\nonumber \\
   &\left.~~~+\left[\binom{N\!-\!3}{n\!-\!3}+\binom{N\!-\!3}{n}\right]E_{n}\right\}
     \bigg/\binom{N}{n}\,,
\end{align}
or
\begin{equation}
\label{En}
(n-1)(E_{n+1}-E_n)=(N-n-1)(E_n-E_{n-1}). 
\end{equation}
\end{subequations}

This second-order recursion can be solved by standard methods~\cite{BO78}, and
the final result is
\begin{subequations}
  \begin{equation}
\label{Pnsol}
E_n=\sum_{j=1}^{n-1} \frac{\Gamma(N-2)}{\Gamma(j)\,\Gamma(N-j-1)}\,,
\end{equation}
where $\Gamma(\cdot)$ is the Euler gamma function.  Correspondingly, the
probability to end with all $\downarrow$ voters is $E_{N-n}$.  In the
continuum limit, this result for $E_n$ simplifies to
\begin{equation}
  E_n\longrightarrow E(y) \simeq
  \tfrac{1}{2}\big[1+\text{erf}(y/\sqrt{2})\big]\,,
\end{equation}
\end{subequations}
with $y=(2n-N)/\sqrt{N}$.  The curve of $E(y)$ versus $y$ is a sigmoidal
function that steepens as $N$ increases and becomes a step function in the
$N\to\infty$ limit.  It becomes extremely unlikely that the initial minority
wins in a large finite population.

\begin{figure}[ht] 
 \vspace*{0.cm}
 \includegraphics*[width=0.375\textwidth]{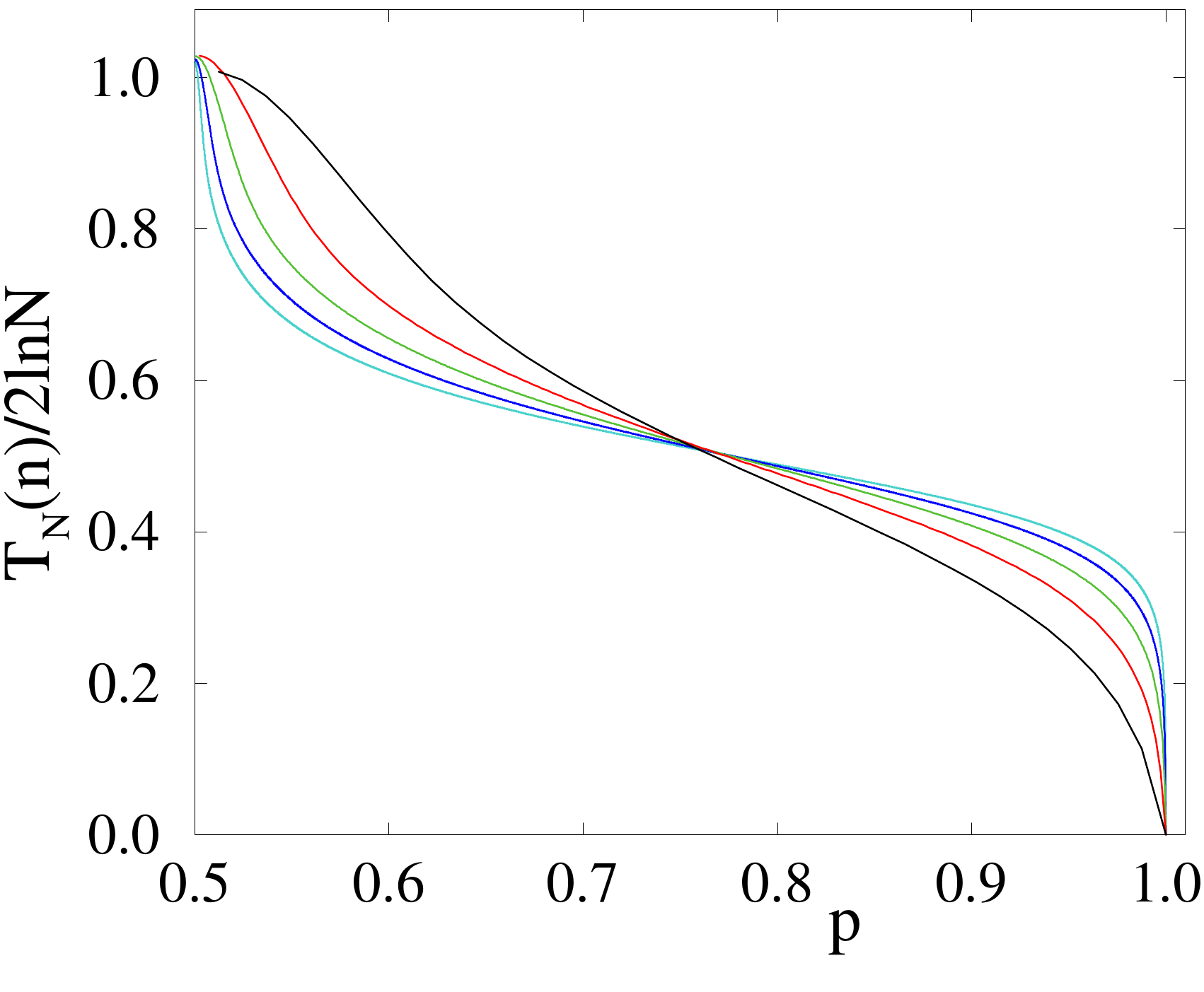}
 \caption{\small Consensus time $T_n$ versus $p=n/N$ for $N=81$, 401, 2001,
   10001, and 50001 (more step-like curves for larger $N$). }
\label{fpt-color}
\end{figure}

Following the same reasoning as above, we also compute the consensus time, $T(n)$,
when starting with $n$ $\uparrow$ voters.  This quantity obeys the recursion
\begin{align}
T_n\! &=\! \left\{3\binom{N\!-\!3}{n\!-\!2} (T_{n+1}\!+\!\delta T)
\!+\!3\binom{N\!-\!3}{n\!-\!1}(T_{n-1}\!+\!\delta T)\right.\nonumber\\
      &\left. ~~+\!\left[\binom{N\!-\!3}{n\!-\!3}+\binom{N\!-\!3}{n}\right]
        (T_{n}\!+\!\delta T)\right\}\bigg/ \binom{N}{n},
\end{align}
subject to the boundary conditions $T(0)=T(N)=0$.  The natural choice for the
time increment of an elemental update is $\delta T=3/N$, so that each spin is
updated once per unit time, on average.  This inhomogeneous second-order
recursion can again be solved by standard methods and the final result is
(Fig.~\ref{fpt-color})
\begin{equation}
\label{Tnsol} 
T_n=1+2k(2k-1)\sum_{j=1}^{n-1}\frac{V_j}{\Gamma(j)\,\Gamma(2k-j)}
\end{equation}
\begin{equation*}
\label{Vjsol}
\hspace{-1.8cm}\text{with}\qquad\qquad
V_j=\sum_{i=1}^{k-j}\frac{\Gamma(k-i)\,\Gamma(k+i-1)}{(k-i+1)\,(k+i)}\,,
\end{equation*}
where $k=\frac{1}{2}(N\!-\!1)$.  While the exact expression is unwieldy, the
$N\to\infty$ behavior of $T_n$ is simple:
\begin{equation}
  T(n)\simeq
  \begin{cases} 2 \ln N &\qquad n=N/2\,,\\
    \ln N & \qquad n\ne N/2\,.
  \end{cases}
\end{equation}

On finite-dimensional lattices, surprisingly rich behavior arises that is
distinct from both the VM and the zero-temperature kinetic Ising model.  In
one dimension, a natural update rule is: select a group of three contiguous
spins that all adopt the majority opinion and repeat this step until
consensus is reached.  At long times, this rule leads to Ising-like dynamics
in which the density of domain walls---the boundary between neighboring
opposite-opinion voters---decays with time as $t^{-1/2}$.

On hypercubic $d$-dimensional lattices, a natural group definition is the von
Neumann neighborhood---a voter and its $2d$ nearest neighbors.  In an update,
voters in a randomly selected neighborhood all adopt the majority opinion,
and this update is repeated until consensus is reached.  This leads to a
dynamics that resembles the kinetic Ising model in that domain interfaces
have non-zero surface tension.  However, straight interfaces, which are
stable in the kinetic Ising model and prevent the ground state from being
reached, are unstable in majority rule.  Thus consensus is always achieved in
majority rule.  Simulations indicate that the consensus time grows as
$N^\nu$, with the exponent $\nu$ continuously decreasing in the spatial
dimension, with numerical values 2, 1.24, 0.72, and 0.56 for spatial
dimension $d=1, 2, 3$, and 4, respectively~\cite{CR05}.  There is currently
no analytical understanding of this unusual dimension dependence of the
consensus time.

\subsection{Nonlinear update rules} 

There are many socially plausible update rules in which the flip rate of a
voter depends nonlinearly on the fraction of disagreeing neighbors.  We
discuss two such examples: (i) the \emph{vacillating} voter model
(VVM)~\cite{LR07} and (ii) the \emph{non-conserved} voter model
(NVM)~\cite{LR08,SSP08}.

In a VVM update, a voter consults two of its neighbors and changes opinion if
it disagrees with either of them.  This irresolution causes a global bias
toward zero magnetization.  Concretely, the VVM update rules are:
\vspace{-2mm}
\begin{enumerate}
\itemsep -1ex
\item A random voter $i$ picks a random neighbor $j$. If $j$ disagrees with
  $i$, then $i$ changes state.
\item If $j$ and $i$ agree, then $i$ picks another random neighbor $k$ and
  adopts its state.
\item  Repeat steps 1--2 until consensus is reached.
\end{enumerate}
It is easily checked that the flip probability for a vacillating voter on the
square lattice equals 0, $\frac{1}{2}$, $\frac{5}{6}$, and 1, respectively,
when the number of misaligned neighbors is 0, 1, 2, and $\geq 3$.  In
contrast, for the VM, the flip probability is $\frac{k}{4}$, where $k$ is the
number of disagreeing neighbors.

In the mean-field limit, the density $x$ of $\uparrow$ voters evolves by the rate
equation
\begin{align}
  \dot x &= -x \left[1-x^2\right] + (1-x)\left[1-(1-x)^2\right]\nonumber\\
  & = x(1-x)(1-2x)\,.
\end{align}
The first term on the right accounts for the loss of $\uparrow$ voters in which a
$\uparrow$ voter is first picked (the factor $x$), and then the neighborhood cannot
consist of two $\uparrow$ voters (the factor $1-x^2$).  Similarly, in the second (gain)
term, a $\downarrow$ voter is first picked, and then the neighborhood must contain at
least one $\uparrow$ voter.  From this rate equation, there are unstable fixed
points at $x=0,1$ and a stable fixed point at $x=1/2$.  Thus the population
is driven to the zero-magnetization state.

\begin{figure}[ht]
\includegraphics[width=0.425\textwidth]{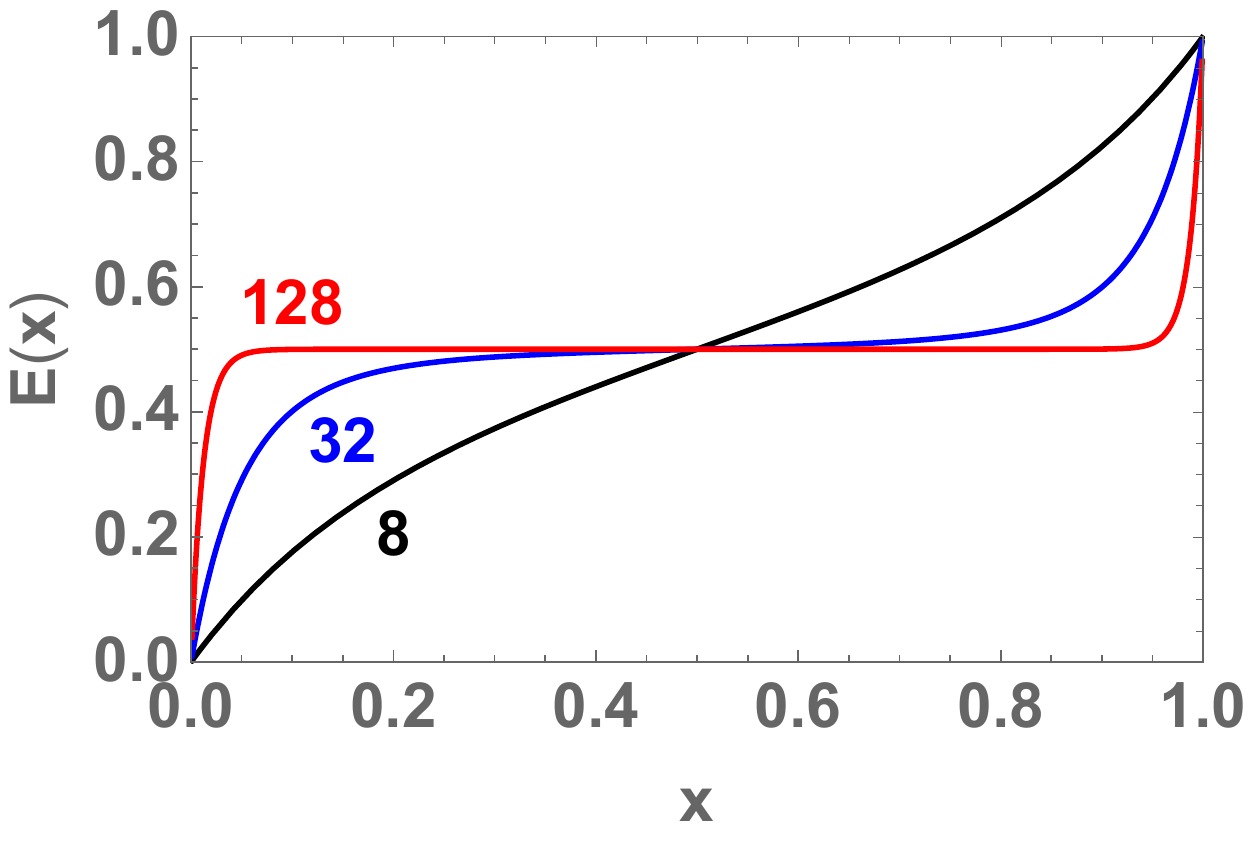}
\vskip - 0.0in
\caption{\small Exit probability $E(x)$ versus the density of $\uparrow$ voters $x$
  for the cases $N=8$, 32, and 128.}
\label{exit}      
\end{figure}

Because consensus is the only absorbing state of the stochastic dynamics, a
finite population ultimately reaches this state.  To characterize the
approach to consensus, we determine the exit probability $E(n)$, the
probability that a population with $n$ $\uparrow$ voters of out $N$ voters
reaches $\uparrow$ consensus.  This exit probability obeys a backward
equation of the form given in Eq.~\eqref{E-def}, with transition
probabilities that are, in the continuum limit,
\begin{align*}
  w_{n\to n+1}&= (1\!-\!x)\left[1-(1\!-\!x)^2\right]\\
  w_{n\to n-1}&= x(1\!-\!x^2)\\
  w_{n\to n}&= x^3+ (1\!-\!x)^3\,.
\end{align*}
Substituting these probabilities in the backward equation and taking the
continuum limit, gives
\begin{equation}
  \label{E-VVM}
  \frac{3 x(1-x)}{2N} \frac{\partial^2E}{\partial x^2} 
  + x(1-x)(1-2x) \frac{\partial E}{\partial x}=0,
\end{equation}
with solution
\begin{equation}
\label{exitMF}
E(x) = \int_{-1/2}^{x-1/2} e^{2Ny^2/3}\, dy\Bigg/ 
\int_{-1/2}^{1/2} e^{2Ny^2/3}\, dy.
\end{equation}
The exit probability approaches an anti-sigmoidal shape in which $E(x)$
attains the nearly constant value $1/2$ over a progressively wider range for
increasing $N$ (Fig.~\ref{exit}).  This behavior reflects the voting bias
towards zero magnetization.  Thus a population with an arbitrary initial
magnetization is driven to an effective potential well at $x=\frac{1}{2}$
(equivalently at $m=0$), and consensus is achieved by the population
surmounting this effective potential barrier.  As a result, the exit
probability becomes nearly independent of $x$ for large $N$.

In the same spirit, the consensus time obeys \eqref{E-VVM} but with the
right-hand side now equal to $-1$.  The formal solution is elementary but
ugly, and is not expressible in closed form.  The main result, however, is
that the consensus time scales exponentially in $N$.  Again, in contrast to
the VM, the global bias drives the population into an effective potential
well that must be surmounted to reach consensus, leading to a large consensus
time.

\begin{figure}[ht]
\includegraphics[width=0.35\textwidth]{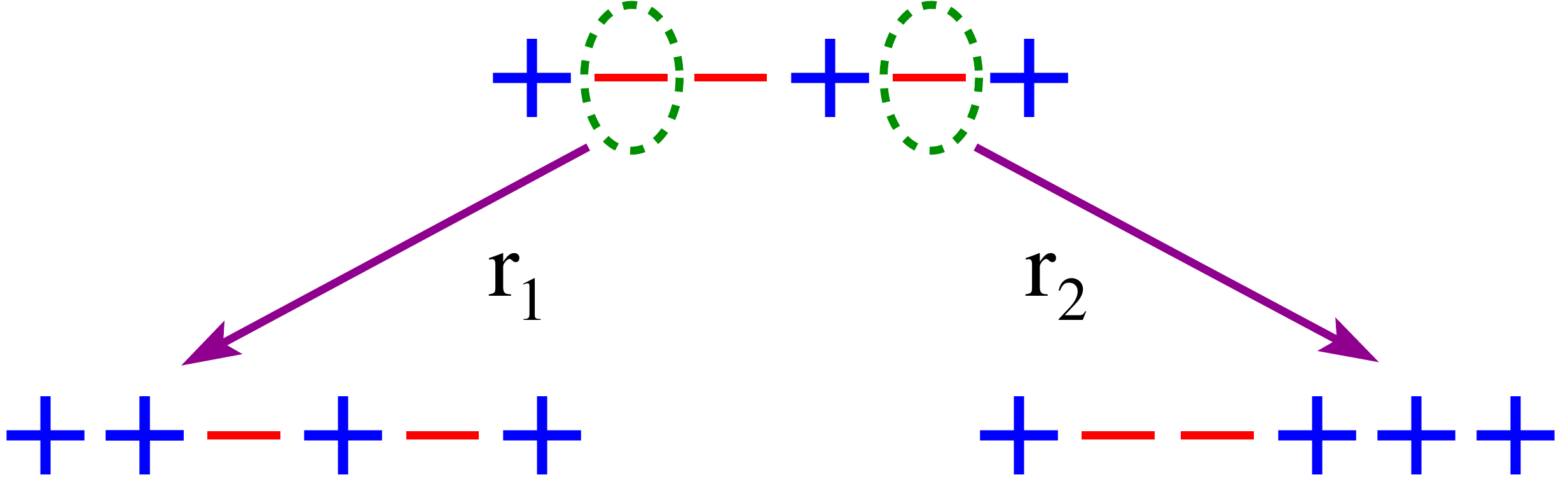}
\caption{\small Update events in the non-conserved voter model in one
  dimension.  A voter changes state with rate $r_1$ if it has 1 disagreeing
  neighbor (left), and with rate $r_2$ if it has 2 disagreeing neighbors
  (right).}
\label{NVM}      
\end{figure}

In the NVM, the flip rate of a voter explicitly depends nonlinearly on the
fraction of disagreeing neighbors.  We treat the simplest setting of one
dimension.  Let $r_f$ denote the flip rate of a voter when the fraction of
disagreeing neighbors is $f$ (Fig.~\ref{NVM}).  It is natural to impose
$r_0=0$, so that no evolution occurs when there is local consensus.  Then the
most general description involves two parameters, $r_1$ and $r_2$.  By
absorbing one rate into the overall time scale, the only parameter is the
ratio $\gamma=r_2/r_1$.  For $\gamma > 2$, the influence of two neighbors is
more than twice that of one neighbor.  For $\gamma\to\infty$, a voter changes
opinion by ``unanimity rule''---all neighbors must have the opposite
opinion~\cite{LTH07}.  The case $\gamma= 2$ corresponds to the classical
voter model.  For $\gamma <2$, one disagreeing neighbor is more effective in
triggering an opinion change than in the classical voter model, and the
limiting case $\gamma=1$ corresponds to the vacillating voter model (VVM).

It is a pleasant game to work out the NVM flip rate in one dimension.  For a
voter at site i, this rate is
\begin{align}
\label{w-NVM}
w_i&\!=\!
\gamma\!+\!2-\gamma\sigma_{i}(\sigma_{i-1}+\gamma\sigma_{i+1})
+  (\gamma\!-\!2 )\sigma_{i-1}\sigma_{i+1}\,.
\end{align}
Left/right symmetry mandates that the $\sigma_{i} \sigma_{i+1}$ and the
$\sigma_{i} \sigma_{i-1}$ terms have the same coefficient.  For $\gamma=2$,
the $\sigma_{i-1}\sigma_{i+1}$ term vanishes, and the VM equation of motion
is recovered.  We shall see, however, that the $\sigma_{i-1}\sigma_{i+1}$
term couples the rate equation for the single-body quantity $\sigma_i$ to
3-body terms (see also Ref.~\cite{G63}).

Because of this coupling, it is not possible to solve the NVM model exactly.
Nevertheless, we obtain an approximate solution by truncating the hierarchy
of rate equations for higher-order correlation functions.  Using
\eqref{dot-sigma}, the time dependence of the mean opinion
$s_i\equiv\langle \sigma_i\rangle$ is given by
\begin{align}
\label{eqS}
\dot s_i = 2\gamma(s_{i+1}\!+\!s_{i-1}) - 2 (\gamma \!+\! 2 )s_i
  - 2 (\gamma \!-\! 2 ) \langle\sigma_{i-1}\sigma_i\sigma_{i+1}\rangle\,.
\end{align}
For $\gamma=2$, the nonlinear term vanishes and the resulting diffusion-like
equation is exactly soluble.  However, for $\gamma\ne 2$, the $s_i$ is
coupled to higher-order correlations.

Let us first neglect all correlations and assume that
$\langle\sigma_{j-1}\sigma_j\sigma_{j+1}\rangle \approx m^3$, with the
magnetization $m= \langle s_i\rangle$.  With this assumption, Eq.~\eqref{eqS}
reduces to
\begin{eqnarray}
\label{mimi}
\dot m  =   2  (\gamma-2) (m-m^3),
\end{eqnarray}
so that the magnetization is not conserved for $\gamma \neq 2$.  The stable
solutions of Eq.~\eqref{mimi} are either consensus ($m=\pm1$) when
$\gamma>2$, or stasis, with equal densities of the two types of voters
($m=0$), when $\gamma <2$.

A better assumption is to to truncate the hierarchy of equations for
multi-spin correlation functions at higher order~\cite{LR07,MR03}.  Consider
the rate equation for the nearest-neighbor correlation function
$\langle\sigma_j \sigma_{j+1}\rangle$:
\begin{align}
\label{ct}
\frac{\partial \langle\sigma_j \sigma_{j\!+\!1}\rangle}{\partial t} &=
- 2 (\gamma-2)\left[\langle\sigma_{j-1} \sigma_{j}\rangle   +
   \langle\sigma_{j+1}\sigma_{j+2}\rangle\right]~~~~\nonumber\\
 &~~~+ 2 \gamma \left[ \langle \sigma_{j-1} \sigma_{j+1} \rangle 
+  \langle \sigma_{j} \sigma_{j+2} \rangle\right]   \nonumber\\[1.5mm]
&~~~+ 4 \gamma  -4 (\gamma + 2 ) \langle \sigma_j \sigma_{j+1} \rangle\, .
\end{align}
To close this equation, we need to approximate the second-neighbor
correlation functions $\langle\sigma_{j} \sigma_{j+2}\rangle$.  To make
progress, let us first examine the role of domain walls---nearest-neighbor
anti-aligned voters---whose density is
$\rho=(1-\langle \sigma_i\sigma_{i+1}\rangle)/2$, on the dynamics.  According
to the transition rate \eqref{w-NVM}, an isolated domain wall diffuses freely
for any $\gamma$.  However, when two domain walls are adjacent, they
annihilate with probability $P_{\rm a}= \gamma/(2+\gamma)$ or they hop away
from each other with probability $P_{\rm h}= 2/(2+\gamma)$.  Thus when
$\gamma>2$, $P_{\rm a}>P_{\rm h}$, and adjacent domain walls tend to
annihilate, while they repel when $\gamma < 2$.  The interaction of two
domain walls is therefore equivalent to single-species annihilation,
$A+A\to 0$, but with a reaction rate that is modified compared to freely
diffusing particles because of this interaction.  Nevertheless, the domain
wall density asymptotically decays as $t^{-1/2}$ for any $\gamma<\infty$, but
with an interaction-independent amplitude \cite{bA97}.

Since domain walls become widely separated at long times, we therefore
approximate the second-neighbor correlation function as
$\langle\sigma_{j} \sigma_{j+2}\rangle \approx \langle\sigma_{j}
\sigma_{j+1}\rangle$~\cite{LR07}.  We also write the nearest-neighbor
correlation function $\langle\sigma_{j} \sigma_{j+1}\rangle$ for a spatially
homogeneous population as $m_2$ for notational simplicity.  Now the rate equation
\eqref{ct} for $m_2$ becomes
\begin{equation}
\label{RE-m2}
\dot m_2 = 4 \gamma - 4 \gamma m_2\,.
\end{equation} 
For the uncorrelated initial condition, $m_2(0)=m(0)^2$, the solution is
\begin{eqnarray}
m_2(t) =  1 + \left[m(0)^2-1\right]\, e^{-4 \gamma t}~,
\end{eqnarray}
where $m(0)\equiv\langle s_j(0)\rangle $ is the magnetization at $t=0$.

In a similar spirit, we decouple the 3-spin correlation function as
$\langle \sigma_{j-1} \sigma_j \sigma_{j+1}\rangle \approx m m_2$ \cite{LR07}
and average over all sites to simplify the rate equation \eqref{eqS} for the
mean spin in a spatially homogeneous population to
\begin{eqnarray}
\dot m  = 2(2-\gamma) me^{-4\gamma t} [m(0)^2-1]\,.
\end{eqnarray}
Solving this equation and taking the $t\to\infty$ limit, we obtain a
non-trivial relation between the final and initial magnetizations:
\begin{align}
\label{m-inf}
m(\infty)  &=  m(0)\, e^{(2-\gamma) (m(0)^2-1)/2\gamma}~ .
\end{align}
 
\begin{figure}[ht]
\includegraphics[angle=0,width=0.4\textwidth]{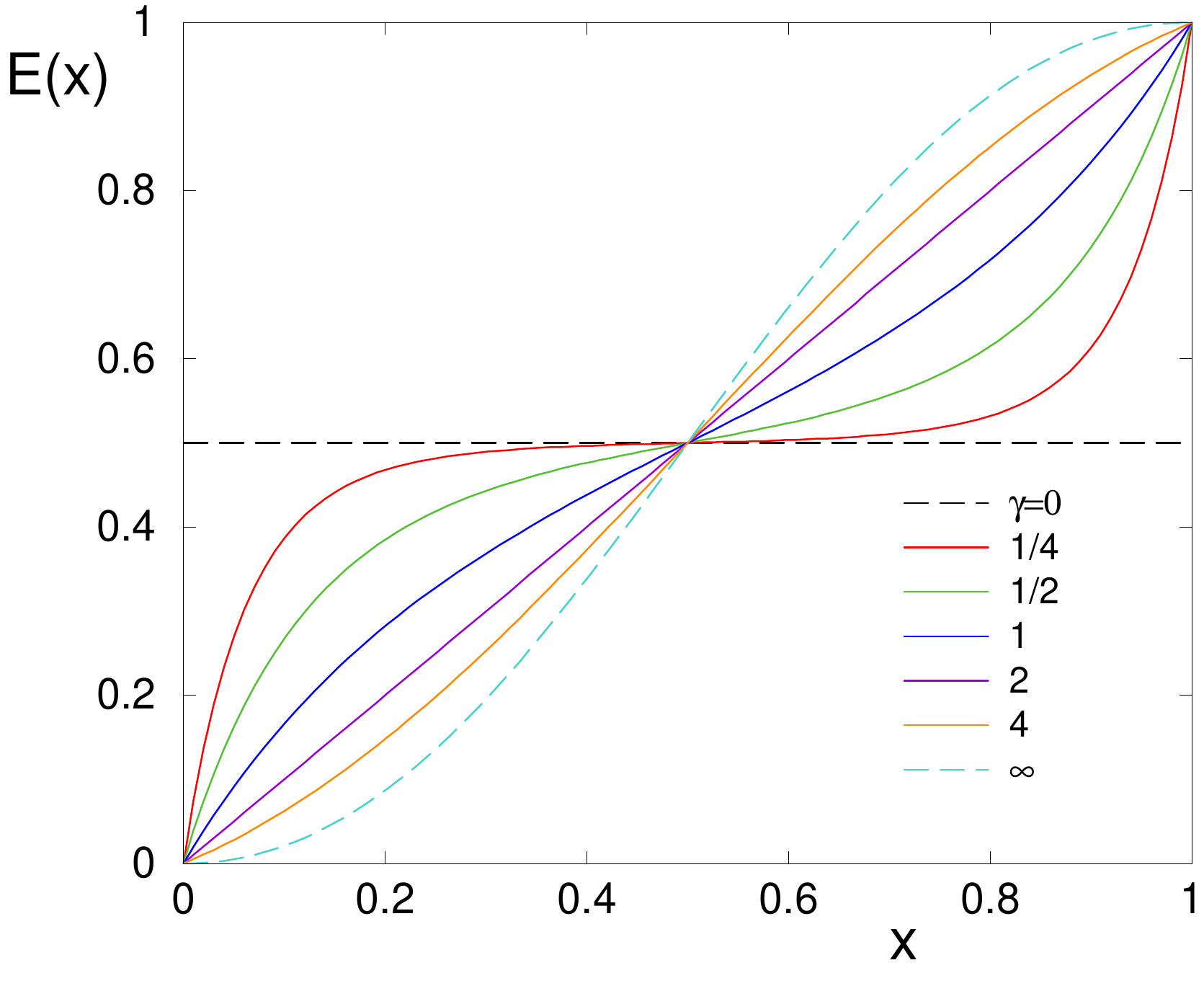}
\caption{\small Exit probability $E(x)$ for the non-conserved voter model in
  one dimension from Eq.~(\ref{exit1D}) as a function of the initial density
  of $\uparrow$ voters $x$ for different values of $\gamma$. }
\label{fig2}      
\end{figure}

However, the average magnetization in a finite population does not
perpetually fluctuate around this asymptotic value but ultimately reaches
$\pm 1$ because consensus is the only absorbing state of the stochastic
dynamics.  We characterize this consensus by the exit probability $E(x)$ to
reach $\uparrow$ consensus when initially $n=x N$ out of $N$ voters are in
the $\uparrow$ state.  Since the density of $\uparrow$ voters $x=(1+m)/2$, and
$m(\infty)= 2E(x)-1$ (see \eqref{m-inf-VM}), Eq.~\eqref{m-inf} gives
(Fig.~\ref{fig2})
 \begin{eqnarray}
 \label{exit1D}
E(x)  =  \frac{1}{2} \left[(2x-1) e^{2x(2-\gamma)(x-1)/\gamma} +1 \right]\,.
\end{eqnarray}
The non-linear behavior of $E(x)$ versus $x$ given Fig.~\ref{fig2} arises
generally in opinion evolution models where the average magnetization is not
conserved, such as the majority rule model of Sec.~\ref{subsec:maj}.  These
sigmoidal (initial slope greater than 1) and anti-sigmoidal curves are
generic behaviors for the exit probability.  Because of the generality of
these two classes, they have been widely investigated in the social science
literature, where a variety of mechanisms have been proposed to generate
these two basic forms for the exit probability~\cite{CW12,ML12}.

\section{More than Two States}

\subsection{Constrained 3-Choice Voting}
\label{subsec:3-choice}

There is nothing sacrosanct about two voting states.  A natural extension of
the VM is to allow a voter to have any number voting states.  For $k$ voting
states and a VM update rule (Sec.~\ref{sec:classic}), the resulting dynamics
is again VM like.  If we label the states as $\{a,b,c,\ldots,k\}$ and
collectively denote the voter states $\{b,c,\ldots,k\}$ as $\overline{a}$,
then the dynamics of $a$ and $\overline{a}$ is just the usual 2-state voter
model, with the initial density of $\overline{a}$ equal to the sum of the
initial densities of $\{b,c,\ldots,k\}$.

\begin{figure}[ht] 
\centerline{ \includegraphics*[width=0.4\textwidth]{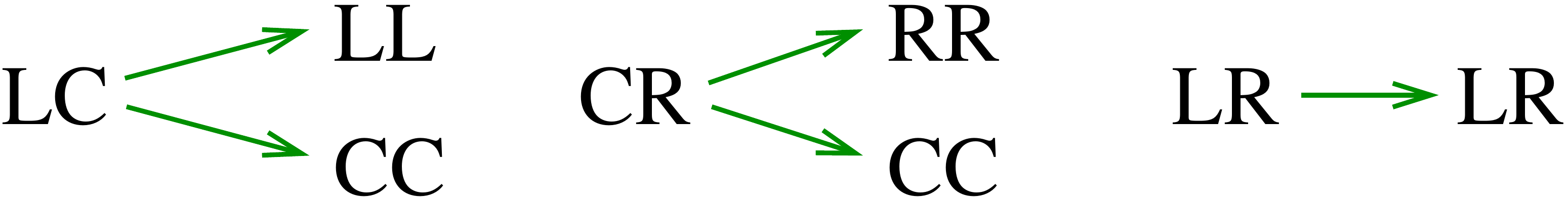}}
 \caption{\small Update events for different pair states in constrained
   3-choice voting.}
\label{3-state-steps}
\end{figure}

More interesting behavior arises if the interactions between the states are
not all identical.  A simple extension of this genre is the \emph{constrained
  3-state voter model}~\cite{VR04} (Fig.~\ref{3-state-steps}), where we
suggestively label the states as leftists $L$ (with density $x$), rightists
$R$ (density $y$), and centrists $C$ (density $z=1-x-y$).  In an elemental
interaction, the pair $LC$ equiprobably transforms to $LL$ or $CC$, while
$CR$ equiprobably transforms to $RR$ or $CC$.  That is, there is usual voter
model dynamics between a centrist and a voter in any other opinion state.
However leftists and rightists are sufficiently separated in opinion space
that do not interact.

In this 3-choice model, updates between interacting pairs occur repeatedly
until the population can no longer evolve.  While the sense of the interactions
are either neutral or consensus promoting, the population can get trapped in
a frozen state that consists of only leftists or consists of only rightists.
This freezing, even though the elemental update is consensus promoting, also
occurs in the venerable Axelrod model~\cite{A77,AAE96,A97} that will be discussed
below.

\begin{figure}[ht] 
 \vspace*{0.cm}
 \includegraphics*[width=0.3\textwidth]{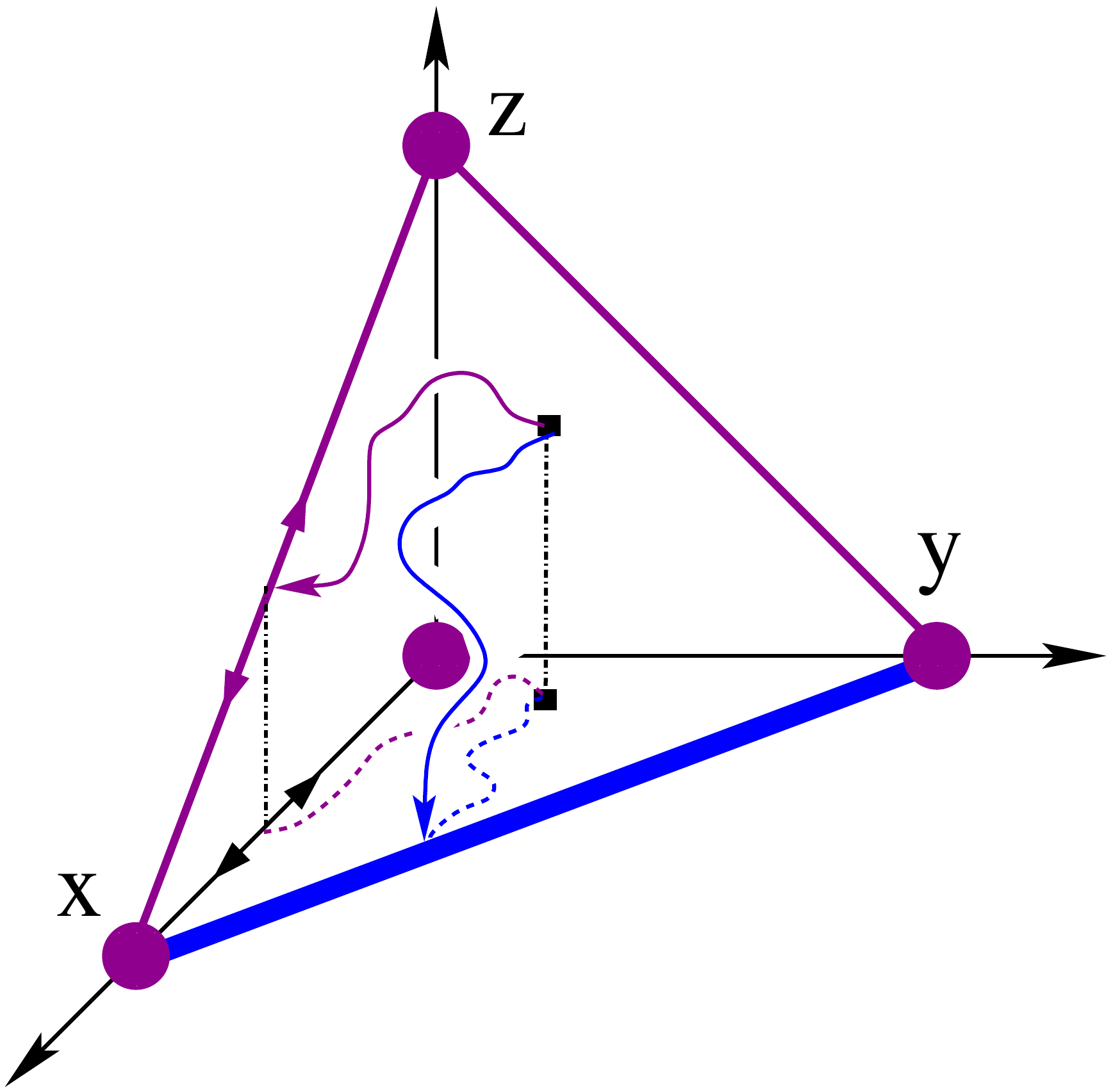}
 \caption{\small The composition triangle $x+y+z=1$.  The purple dots denote
   consensus states and the heavy blue line denotes frozen final states where
   no centrists remain.  Typical trajectories are shown.  When a trajectory
   reaches the lines $x=0$ or $y=0$, the trajectory subsequently remains on
   this line until consensus is necessarily reached. }
\label{triangle}
\end{figure}

We now outline how to determine the probability $F(x,y)$ that the population
reaches a frozen state as a function of the initial densities, $x$ and $y$.
This quantity is just the first-passage probability for a trajectory to hit
the line $x+y=1$ when it starts at some point within the tetrahedral region
$x+y+z<1$ in the composition space spanned by the densities $x,y,$ and $z$
(Fig.~\ref{triangle}).  We need to impose the boundary conditions $F=0$ for
$x=0$ and $y=0$, where the probability of reaching the frozen state is zero,
and $F=1$ on the line segment $x+y=1$.  This first-passage probability obeys
the backward Kolmogorov equation (see the discussion that explained
Eq.~\eqref{E-def})
\begin{align}
\label{Fxy}
\begin{split}
F(x,y)&= p_x[F(x-\delta,y)+F(x+\delta,y)] \\
 &+ p_y[F(x,y-\delta)+F(x,y+\delta)]  \\
 &+ [1-2(p_x+p_y)]F(x,y)\,,
\end{split}
\end{align}
where
\begin{align*}
  p_x&={N_-N_0}/[N(N-1)]\\
  p_y&={N_+N_0}/[N(N-1)]
\end{align*}
are the transition probabilities for $(N_-,N_0)\to(N_-\pm 1,N_0\mp 1)$ and
$(N_+,N_0)\to(N_+\pm 1,N_0\mp 1)$, respectively, and $\delta=1/N$.
  
\begin{figure}[ht] 
 \vspace*{0.cm}
 \includegraphics*[width=0.3\textwidth]{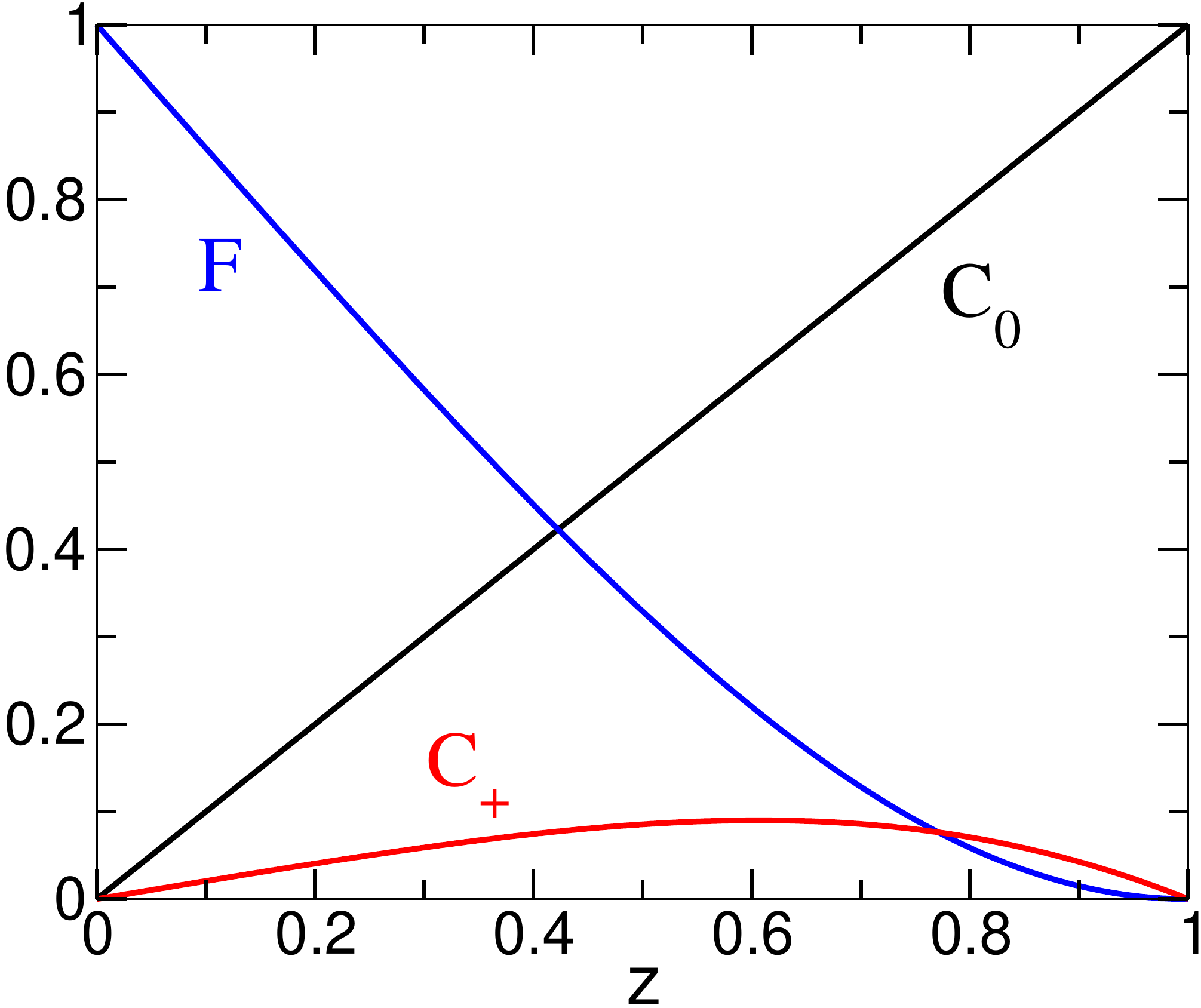}
 \caption{\small The probabilities to reach the frozen state, $F$, and to
   reach the centrist and extremist consensus states, $C_0$ and $C_+$, respectively,
   as a function of the initial density of centrists $z$ for the case of
   equal initial densities of leftists and rightists, $y/x=1$.}
\label{solution}
\end{figure}

In the continuum limit, \eqref{Fxy} reduces to
\begin{equation}
 x\,\frac{\partial^2 F(x,y)}{\partial x^2}+y\,\frac{\partial^2 F(x,y)}
 {\partial y^2}=0\,,
 \label{FP_eq}
\end{equation}
subject to the mixed boundary conditions $F(x,0)=F(0,y)=0$, $F(x,1-x)=1$.
This mixed boundary condition is a complicating feature that seems to prevent
a direct solution.  However, after a series of coordinate transformations,
Eq.~\eqref{FP_eq} can be mapped onto a Schr\"odinger equation in the presence
of a specific potential well, which is soluble~\cite{VR04}.  By these means,
the solution to Eq.~\eqref{FP_eq} is~\cite{VR04}
\begin{equation}
 F(x,y)=\!\!\sum_{n~ {\rm odd}} \frac{2(2n+1)}{n(n+1)} \, \sqrt{xy}\, (x\!+\!y)^n \, 
 P_n^1\left(\frac{x\!-\!y}{x\!+\!y}\right),
 \label{sol1}
\end{equation}
where $P_n^1$ is the associated Legendre function.  By a similarly tedious
calculation, the probability to reach $\downarrow$ consensus as a function of the
initial densities $x$ and $y$ is
\begin{equation}
 C_-(x,y)
 \label{P+1}
 =x-\sum_{n=1}^\infty \frac{(2n+1)}{n(n+1)} \, \sqrt{xy}\, (x\!+\!y)^n \,  
 P_n^1\left(\frac{x\!-\!y}{x\!+\!y}\right)
\end{equation}
The probability to reach $\uparrow$ consensus, $C_+(x,y)$, is obtained by the
obvious symmetry $C_+(x,y)=C_-(y,x)$.

One of the main features of the final-state probabilities in
Fig.~\ref{solution} is that centrists are needed to catalyze consensus.  As
the centrist density $z\to 1$, the probability of reaching a frozen state
goes to zero.  Another comforting feature of this figure is that extremist
consensus is relatively unlikely to occur.

\subsection{Axelrod Model}

Some societies culturally fragment, even though individuals may try to reach
agreement with acquaintances.  The Axelrod model provides a simple
description for this dichotomy~\cite{A77,AAE96,A97}.  Here, each individual
possesses $F$ characteristic features---for example, preferences for sports,
for music, for food, {\it etc}---that each can assume $q$ distinct values.
In an update, a pair of agents is selected at random.  If the agents disagree
on all features, no interaction occurs.  However, if the agents agree on at
least one feature, they interact with probability equal to the fraction of
shared features.  In an interaction, a previously unshared feature is
selected at random and one agent copies the preference for this feature from
its interaction partner.  While the interaction ostensibly brings agents
closer together, the restriction that only sufficiently compatible
individuals can interact prevents this convergence from reaching consensus.

Depending on the parameters $(F,q)$, the Axelrod model on a
finite-dimensional lattice undergoes a phase transition between consensus and
a frozen discordant state, where each interacting pair is incompatible
\cite{A77,AAE96,A97,CMV00,VVC02,KETM03a,KETM03b}.  In the mean-field limit,
the corresponding transition is between a steady state with perpetual social
``churn'' and a frozen state.  There is also a non-monotonic time dependence
of the societal activity level~\cite{CMV00} that occurs over a very long time
scale~\cite{CMV00,VR07}.  This long time scale is unexpected because the
underlying dynamical equations (Eqs.~\eqref{eqn-PA} below) have rates of the
order of one.  Important examples where dynamics with rates of the order of 1
leads to anomalously long time scales include the Lorenz model~\cite{L63},
the three-species competition species models~\cite{ML75}, and
HIV~\cite{PN99}.  In the Lorenz model, although the dynamics is a contracting
map, trajectories can fall into a strange attractor with very long period
dynamics.  In HIV, after an individual contracts the disease, there is an
immune response over several months, followed by a latency period that can
last beyond a decade, during which an individual's T-cell level slowly
diminishes.  Our results for the Axelrod model provide some insight into how
widely separated time scales can arise in a simple dynamical system.

To simplify the modeling as much as possible, we discuss a mean-field
version in which each agent has a small and fixed number $z$ of interaction
partners, corresponding to a degree-regular random graph.  We also restrict
to the simplest non-trivial case of $F=2$ features.  Thus there are 3 types
of links between individuals: links with no shared features (type 0), and
links with both features shared (type 2).  These two types of links are
inactive because neither individual changes its state across such links.
Finally, there are links of type 1 where a single feature is shared, which
are the only active links in the population.

By accounting for the various interaction events, the time dependence of the
fraction of links of type $i$ ($i=0,1,2$) when a single agent changes its
state is described by the rate equations~\cite{VR07}
\begin{align}
\label{eqn-PA}
\begin{split}
\dot P_0 &\!=\!\frac{z-1}{z} P_1 \left( -\lambda P_0 + 
\frac{1}{2} P_1 \right)\,, \\
\dot P_1&\!=\!\!-\frac{P_1}{z}+\frac{z-1}{z} P_1\! 
\left( \lambda P_0 -      \frac{1\!+\!\lambda}{2} P_1 + P_2 \right), \\
\dot P_2&\!=\!\frac{P_1}{z}+\frac{z-1}{z} P_1 \left( \frac{\lambda}{2}
  P_1 - P_2 \right)\,,
\end{split}
\end{align}
where $z$ is the number of neighbors of each individual.  The first term on
the right-hand sides of the first two equations account for the direct
interaction between agents $i$ and $j$ that changes a link of type $1$ to
type $2$.  For example, in $\dot P_1$, a type-1 link and the shared feature
across this link is chosen with probability $P_1/2$ in an update event.  This
update decrements the number of type-1 links by one in a time $dt={1}/{N}$,
where $N$ is the population size.  Assembling these factors gives the term
$-{P_1}/{z}$ in this equation.

\begin{figure}[ht]
\includegraphics[width=0.325\textwidth]{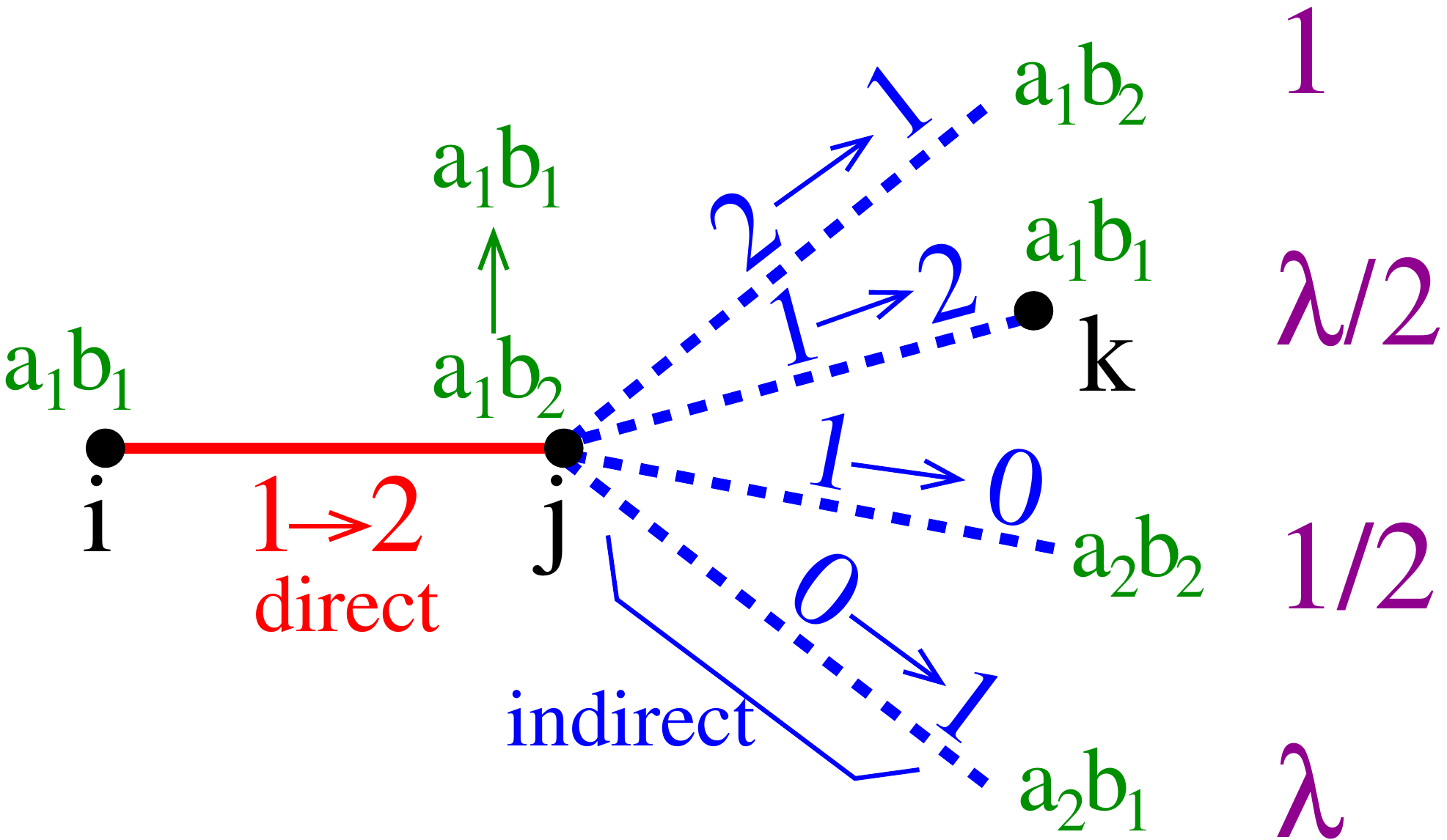}
\caption{\small State-changing updates on link $jk$ when agent $j$ changes state
  from $a_1b_2\to a_1b_1$ in an interaction with $i$.  The values at the
  right give the relative rates of each type of event.}
\label{states}
\end{figure}

The remaining terms in the rate equations represent indirect interactions.
For example, if agent $j$ changes state from $(a_1,b_2)$ to $(a_1,b_1)$ then
the link $jk$ that joins to agent $k$ in state $(a_1,b_1)$ changes from type
1 to type 2 (Fig.~\ref{states}).  The probability for this event is
proportional to $\lambda P_1/2$: $P_1$ is the probability that the indirect
link is of type 1, the factor 1/2 accounts for the fact that only the first
feature of agents $j$ and $k$ can be shared, while $\lambda$ is the
conditional probability that $i$ and $k$ share one feature that is
simultaneously not shared with $j$.

\begin{figure}
  \includegraphics[width=0.4\textwidth]{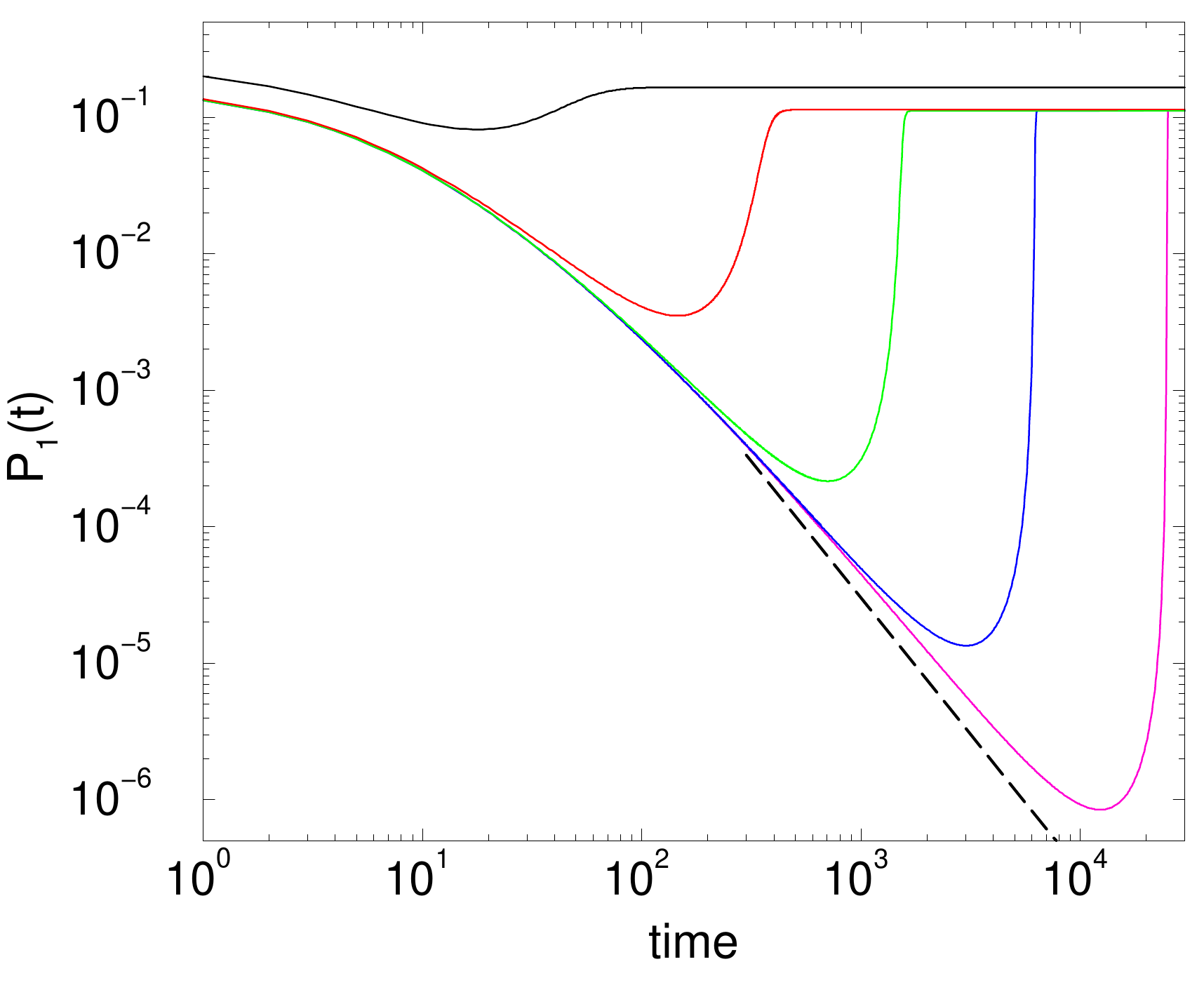}
  \caption{\small Density of active links $P_1(t)$ for $q=q_c-4^{-k}$, with
    $k=-1, 1, 3, 5$, and $7$ (progressively lower minima).  Each agent has
    $z=4$ neighbors.  The dashed curve shows how $P_1\to 0$ for
    $q=q_c+4^{-6}$. }
\label{P1}
\end{figure}

If the distribution of preferences is uniform, then $\lambda = 1/(q-1)$.
While $\lambda$ generally depends on the densities $P_m$, the simulations
presented in Ref.~\cite{VR07} indicate that $\lambda$ is roughly constant and
close to $1/(q-1)$.  When we use this assumption, the rate
equations~\eqref{eqn-PA} become soluble.  The solution itself is too unwieldy
to display here (the full solution is given in Ref.~\cite{VR07}), but the
main result is quite striking (Fig.~\ref{P1}).  The crucial feature is that
there is a dramatic non-monotonicity in the density of active links, in which
the time scale for the crossover between a nearly inactive state and an
active steady state diverges as $q\to q_c=2(z-1)+2\sqrt{(z-1)(z-2)}$.  While
this strange behavior is the prediction from the mathematical solution to the
rate equations~\eqref{eqn-PA}, the physical mechanism that underlies this
long-time non-monotonicity is not understood.

\section{Compromise Models}

\begin{figure}[ht] 
 \vspace*{0.cm}
\vskip 0.05in
 \includegraphics*[width=0.4\textwidth]{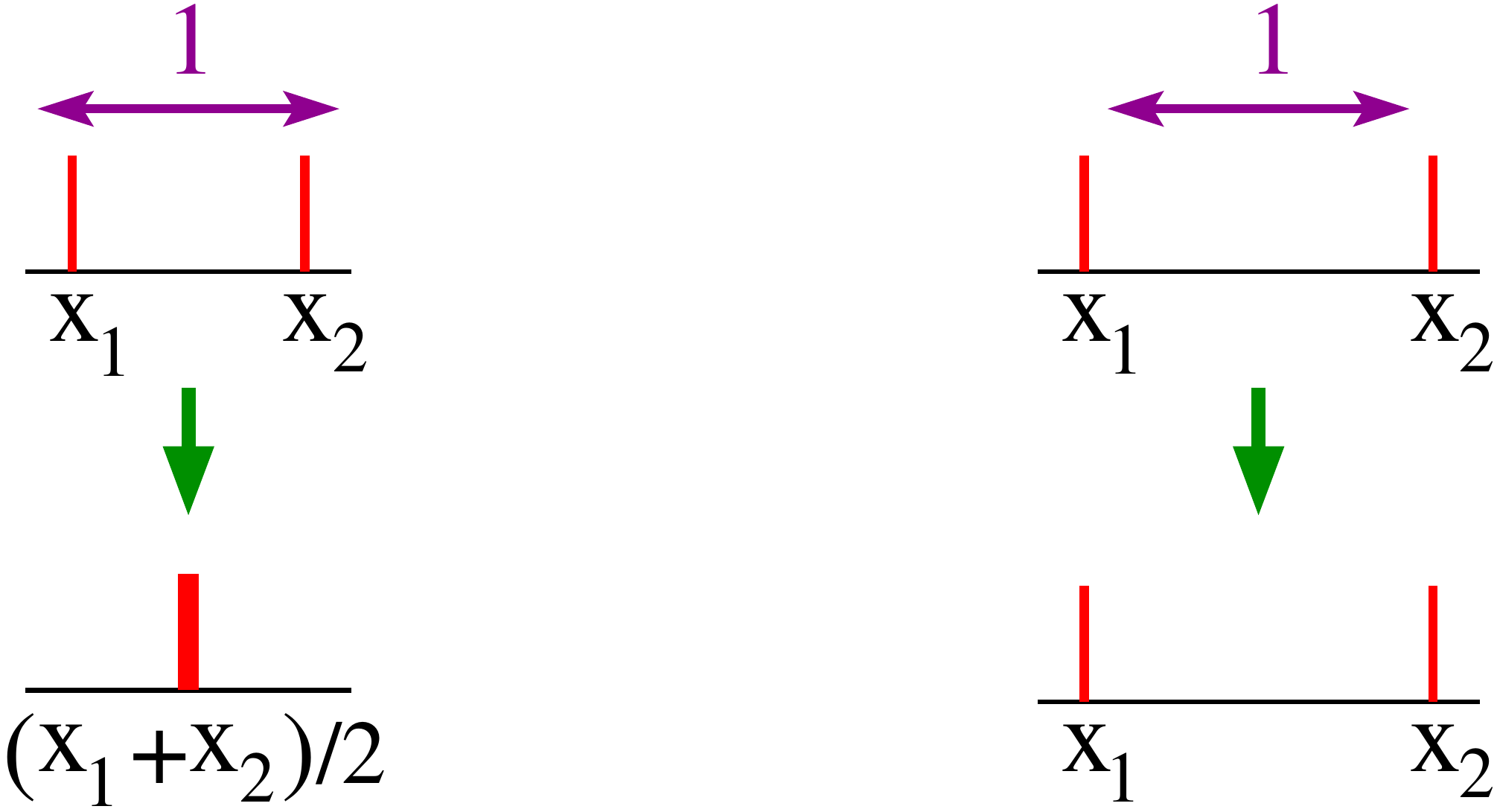}
 \caption{\small Possible updates in the compromise model.  Two agents with
   opinions $x_1$ and $x_2$ with $|x_2-x_1|<1$ compromise.  Otherwise there
   is no opinion change.}
\label{compromise-def}
\end{figure}

The bounded confidence, or compromise, model accounts for the human attribute
that individuals often compromise their positions in an interaction.  This
class of appealing models was introduced in the physics literature in
Refs.~\cite{WDAN02,HK02}.  For simplicity, we take the opinion of individual
$i$ as a real coordinate $x_i$ in a one-dimensional opinion space.  The state
of the population is updated as follows: Two agents are selected at random.
If their opinions are separated by a distance less than a threshold that we
take as fixed and equal to 1, they compromise; conversely, if their opinions
are separated by a distance larger than 1, they do not interact
(Fig.~\ref{compromise-def}).  This update step is repeated until the opinions
of all individuals reach a static limit.

Suppose that each agent is initially assigned an opinion $x$ from the uniform
distribution in $[-\Delta,\Delta]$, with $\Delta$ the only model parameter.
According to the update rule given above, the opinions of a randomly selected
pair of agents change according to (Fig.~\ref{compromise-def})
\begin{align*}
(x_1,x_2)&\to \tfrac{1}{2}\left(x_1+x_2,x_1+x_2\right) &\qquad |x_2-x_1|<1\,,\\
(x_1,x_2)&\to (x_1,x_2) &\qquad |x_2-x_1|>1\,.
\end{align*}
A natural way to determine the time evolution of the distribution of
opinions, $P(x,t)$, is to perform numerical simulations of the process
defined above.  However, such simulations become inefficient when agents
opinions are close.  In this case, it becomes necessary to update the opinions of
agents many times for their opinions to change by a small amount.

It is much more efficient to numerically integrate the master equation that
describes the time dependence of $P(x,t)$ because there is no ``slowing
down'' of the dynamics.  The master equation for the opinion distribution is
\begin{align}
\label{pxt}
\frac{\partial}{\partial t}P(x,t)&=\!\!\!\!\iint\limits_{|x_1-x_2|<1}\!\!\!\!
dx_1dx_2P(x_1,t)P(x_2,t)\nonumber\\
&~~\times\left[\delta\left((x-\tfrac{1}{2}(x_1\!+\!x_2)\right)-\delta(x\!-\!x_1)\right].
\end{align}

\begin{figure}[ht] 
 \vspace*{0.cm}
\vskip 0.05in
 \includegraphics*[width=0.425\textwidth]{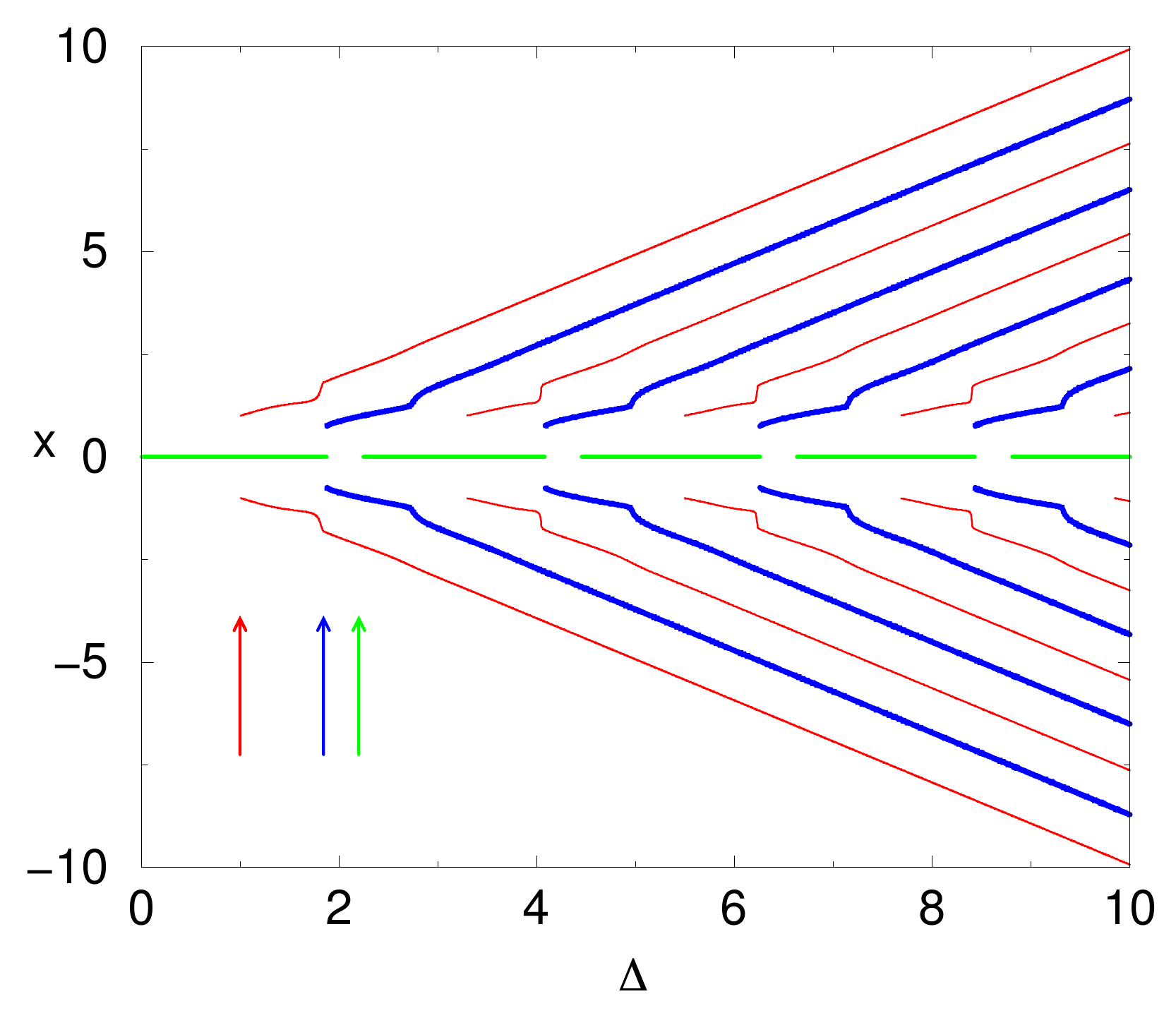}
 \caption{\small The final opinion distribution as a function of the
   interaction threshold $\Delta$.  Vertical arrows mark the first 3
   bifurcations at $\Delta_1=1$, $\Delta_2\approx 1.871$ and
   $\Delta_3\approx 2.248$.}
\label{bifurcation}
\end{figure}

While this equation does not appear to be soluble, it can be integrated with
high precision~\cite{BKR03} in an efficient way, and the strikingly beautiful
final state bifurcation diagram is shown in Fig.~\ref{bifurcation}.  For
$\Delta<\Delta_1=1$, all pairs of individuals are compatible and the final
state is consensus at $x=0$.  However, for $\Delta>\Delta_1$, a bifurcation
arises in which a small fraction of extremists---both positive and
negative---splits off from centrist consensus and forms a positive and a
negative extremist group.  As $\Delta$ increases still further, the extremists
become progressively more extreme until a second bifurcation at
$\Delta=\Delta_2\approx 1.871$, where the centrist group splits into
center-left and center-right groups, with nobody left in the center.  At
$\Delta=\Delta_3\approx 2.248$, there is a third bifurcation where leftists
and rightists have moved sufficiently far from the center that a new centrist
group can nucleate.

As $\Delta$ continues to increase, this sequence of bifurcations
systematically repeats.  For fixed $\Delta$, the spectrum of opinion states
qualitatively mirrors what has happened in some multiparty parliamentary
democracies.  The important feature of the compromise model is that a too
narrow compromise range leads to a fragmented polity, where each party lives
within its own ``echo chamber'' and has no interaction with other parties.

\section{OUTLOOK}

The venerable voter model has played a central role in probability theory and
in statistical physics because it is one of the few exactly soluble
many-particle interacting systems.  The voter model has also been a starting
point to describe a variety of social phenomena.  However, the basic voter
model is clearly too idealized to be of direct empirical relevance and much
research has been devoted to incorporating socially motivated aspects of
decision making into the model.  In the absence of a correspondence between
model parameters and empirical data, these generalizations should not be
oversold as descriptions of social reality, but rather, as potentially useful
descriptions of how opinions in a large population can change over time.

In this short review, a number of these extensions of the voter model were
presented.  These generalizations by no means cover the range of work of this
genre, but we hope to have given the reader a useful perspective.  In spite
of their obvious shortcomings, each of the models presented here reveals rich
phenomenology and appealing methodological aspects.

A basic message from these modeling efforts is that incorporating any
realistic feature of decision making typically leads to either a dramatically
hindered approach to consensus or to the prevention of consensus altogether.
This prevention of consensus addresses one of the glaring unrealistic
features of the voter model in that consensus is always achieved.  Models
such as those outlined here can also help to quantify verbal predictions that
often appear in the social science literature and help determine which types
of models could reproduce empirical observations.  Thus from the optimistic
perspective, perhaps some of these generalizations point the way to properly
calibrated models of social dynamics.

I thank my collaborators who contributed substantially to the results
presented here: Tibor Antal, Eli Ben-Naim, Pu Chen, Nathaniel Gibert, Paul
Krapivsky, Renaud Lambiotte, Naoki Masuda, Vishal Sood, Federico Vazquez, and
Dan Volovik.  I am also grateful to Mirta Galesic for a critically reading of
a draft of this manuscript and for many helpful suggestions.  I also thank
the NSF for financial support during the period of this research through
grants DMR-0227670, DMR-0535503, DMR-0906504, as well as DOE grant
W-7405-ENG-36 when the author spent a year at the Center for Nonlinear
Studies at Los Alamos National Lab.


\newpage

\begin{thebibliography}{99}

\bibitem{CS73} P. Clifford and A. Sudbury, ``A Model for Spatial Conflict'',
  Biometrika \textbf{60}, 581 (1973).

\bibitem{HL75} R. A. Holley and T. M. Liggett. `` Ergodic Theorems for Weakly
  Interacting Infinite Systems and the Voter Model'', Cox, Ann.\ Probab.\
  \textbf{3}, 643 (1975).

\bibitem{C89} J. T. Cox, ``Coalescing Random Walks and Voter Model Consensus
  Times on the Torus in $\mathbb{Z}$'', Ann.\ Probab.\ \textbf{17}, 1333
  (1989).

\bibitem{L99} T. M. Liggett, {\it Stochastic Interacting Systems: Contact,
    Voter, and Exclusion Processes} (Springer, New York, 1999).

\bibitem{K92} P. L. Krapivsky, ``Kinetics of Monomer-Monomer Surface
  Catalytic Reactions'', Phys.\ Rev.\ A \textbf{45}, 1067 (1992).

\bibitem{FK96} L.~Frachebourg and P.~L.~Krapivsky, ``Exact Results for
  Kinetics of Catalytic Reactions'', Phys.\ Rev.\ E {\bf 53},
  R3009 (1996).

\bibitem{DCCH01} I. Dornic, H. Chat\'e, J. Chave, and H. Hinrichsen,
  ``Critical Coarsening Without Surface Tension: The Universality Class of
  the Voter Model'', Phys.\ Rev.\ Lett.\ {\bf 87}, 045701 (2001).
 
\bibitem{CFL09} C. Castellano, S. Fortunato, and V. Loreto, ``Statistical
  Physics of Social Dynamics'', Rev.\ Mod.\ Phys.\ \textbf{81}, 591 (2009).
  
\bibitem{KRB10} P. L. Krapivsky, S. Redner, and E. Ben-Naim, \emph{A Kinetic
    View of Statistical Physics} (Cambridge University Press, Cambridge, UK,
  2010).

\bibitem{B18} A. Baronchelli, ``The Emergence of Consensus: A Primer'',
  Roy.\ Soc.\ Open Sci.\ \textbf{5}, 172189 (2018).
  
\bibitem{JS19} A. J\c{e}drzejewski and K Sznajd-Weron, ``Statistical Physics Of
  Opinion Formation: is it a SPOOF?'' (2019).
  
\bibitem{FFF99} M. Faloutsos, P., Faloutsos, and C. Faloutsos, ``On Power-Law
  Relationships of the Internet Topology'', Comput.\ Commun.\ Rev.\
  \textbf{29}, 251 (1999).

\bibitem{BKM00} A. Broder, R. Kumar, F. Maghoul, P. Raghavan, S. Rajagopalan,
  R. Stata, A.  Tomkins, and J. Wiener, ``Graph Structure in the Web'',
  Comput.\ Networks \textbf{33}, 309 (2000).

\bibitem{N01} M. E. J. Newman, ``The Structure of Scientific Collaboration
  Networks'', Proc.\ Natl.\ Acad.\ Sci.\ USA \textbf{98}, 404 (2001).

\bibitem{GDB06} T. Gross, C. J. D. D'Lima and B. Blasius, ``Epidemic Dynamics
  on an Adaptive Network'', Phys.\ Rev.\ Lett.\ \textbf{96}, 208701 (2006).
  
\bibitem{HN06} P. Holme and M. E. J. Newman, ``Nonequilibrium Phase
  Transition in the Coevolution of Networks and Opinions'', Phys.\ Rev.\ E
  \textbf{74}, 056108 (2006).

\bibitem{KB08} B. Kozma and A. Barrat, ``Consensus Formation on Adaptive
  Networks'', Phys.\ Rev.\ E \textbf{77}, 016102 (2008).

\bibitem{SS08} L. B. Shaw and I. B. Schwartz, ``Fluctuating Epidemics on
  Adaptive Networks'', Phys.\ Rev.\ E \textbf{77}, 066101 (2008).

\bibitem{SS10} L. B. Shaw and I. B. Schwartz, ``Enhanced Vaccine Control of
  Epidemics in Adaptive Networks'', Phys.\ Rev.\ E \textbf{81}, 046120 (2010)

\bibitem{DGL12} R. Durrett, J. P. Gleeson, A. L. Lloyd, P. J. Mucha, F. Shi,
  D. Sivakoff, J. E. S. Socolar, and C. Varghese, ``Graph Fission in an
  Evolving Voter Model'', Proc.\ Natl.\ Acad. (USA) \textbf{109}, 3682
  (2012).

\bibitem{RG13} T. C. Rogers and T. Gross, ``Consensus Time and Conformity in
  the Adaptive Voter Model'', Phys.\ Rev.\ E \textbf{88}, 030102 (2013).

\bibitem{GS18} M. Galesic and D. L. Stein, ``Statistical Physics Models of
  Belief Dynamics: Theory and Empirical Tests'', Physica A \textbf{519}, 275
  (2019).
  
\bibitem{GSS83} J.~D.~Gunton, M.~San~Miguel, and P.~S.~Sahni, in: {\it Phase
    Transitions and Critical Phenomena}, Vol.~8, eds.\ C.~Domb and
  J.~L.~Lebowitz (Academic, NY 1983).

\bibitem{B94} A.~J.~Bray, ``Theory of Phase-Ordering Kinetics'', Adv.\ Phys.\
  {\bf 43}, 357 (1994).

\bibitem{K31} A. Kolmogoroff, ``On Analytical Methods in Probability
  Theory'', Math.\ Ann.\ \textbf{104} 415 (1931).

\bibitem{K97} N. G. van Kampen, \emph{Stochastic Processes in Physics and
    Chemistry}, $2^{\rm nd}$ ed.\ (North-Holland, Amsterdam, 1997).

\bibitem{R01} S. Redner, \emph{A Guide to First-Passage Processes}, (Cambridge
  University Press, New York, 2001).

\bibitem{A51} S. E. Asch, in \emph{Groups, Leadership and Men}, ed.\
  H. Guetzkow (Carnegie Press, Pittsburgh, PA 1951).

\bibitem{KBR18} R. L. Kendal, N. J. Boogert, L. Rendell, K. N. Laland,
  M. Webster, and P. L. Jones, ``Social Learning Strategies: Bridge-Building
  Between Fields'', Trends in Cog.\ Sci.\ \textbf{22}, 651 (2018).
  
\bibitem{MGR10} N. Masuda, N. Gibert, and S. Redner, ``HeterogeneouS Voter
  Models'', Phys.\ Rev.\ E \textbf{82}, 010103 (2010).

\bibitem{G78} M. Granovetter, ``Threshold Models of Collective Behavior'',
  Am.\ J. Sociol.\ \textbf{83}, 1420 (1978).

\bibitem{W02} D. J. Watts, ``A Simple Model of Global Cascades on Random
  Networks'', Proc.\ Natl.\ Acad.\ Sci.\ (USA) \textbf{99}, 5766 (2002).

\bibitem{M08} M. O. Jackson, \emph{Social and Economic Networks} (Princeton
  University Press Princeton, NJ, 2008).
  
\bibitem{G87} J. Galambos, {\it The Asymptotic Theory of Extreme Order
    Statistics} (Krieger Publishing Co., Malabar, FL, 1987).

\bibitem{M80} S. Moscovici, ``Toward A Theory of Conversion Behavior'', Adv.\
  Experimental Soc.\ Psych.\ \textbf{13}, 209 (1980).

\bibitem{M85} S. Moscovici, ``Innovation and Minority Influence'', in
  \emph{Perspectives on Minority Influence} Eds.\ S. Moscovic, G Mugny, and
  E. Van Vermaet (Cambridge University Press, Cambridge, UK, 1985).

\bibitem{GJ07} S. Galam and F. Jacobs, ``The Role of Inflexible Minorities in
  the Breaking of Democratic Opinion Dynamics'', Physica A \textbf{381}, 366
  (2007).
  
\bibitem{XSK11} J. Xie, S. Sreenivasan, G. Korniss, W. Zhang, C. Lim, and
  B. K. Szymanski, ``Social Consensus Through the Influence of Committed
  Minorities'', Phys.\ Rev.\ E, \textbf{84}, 011130 (2011).
  
\bibitem{C10} D. Centola, ``The Spread of Behavior in an Online Social
  Network Experiment'', Science \textbf{329}, 1194 (2010).
 
\bibitem{CMP09} C. Castellano, M.A. Mu\~noz, and R. Pastor-Satorras,
  ``Nonlinear q-Voter Model'' Phys.\ Rev.\ E \textbf{80}, 041129 (2009).

\bibitem{DW04} P. S. Dodds and D. J. Watts, ``Universal Behavior in a
  Generalized Model of Contagion'', Phys.\ Rev.\ Lett.\ \textbf{92}, 218701
  (2004).

\bibitem{VR12} D. Volovik and S. Redner, ``Dynamics of Confident Voting'',
  J. Stat.\ Mech.\ P04003 (2012).

\bibitem{SEM04} K. Suchecki, V. M. Equ\'iluz, and M. San Miguel,
  ``Conservation Laws for the Voter Model in Complex Networks'', Europhys.\
  Lett.\ \textbf{69}, 228 (2004).

\bibitem{SEM05} K. Suchecki, V. M. Equ\'iluz, and M. San Miguel, ``Voter
  Model Dynamics in Complex Networks: Role of Dimensionality, Disorder, and
  Degree Distribution'', Phys.\ Rev.\ E \textbf{72}, 036132 (2005).

\bibitem{CLB05} C. Castellano, V. Loreto, A. Barrat, F.  Cecconi, and
  D. Parisi, ``Comparison of Voter and Glauber Ordering Dynamics on
  Networks'', Phys.\ Rev.\ E \textbf{71}, 066107 (2005).

\bibitem{SR05} V. Sood and S. Redner, ``Voter Model on Heterogeneous
  Graphs'', Phys.\ Rev.\ Lett.\ {\bf 94}, 178701 (2005).

\bibitem{ARS06} T. Antal, S. Redner, and V. Sood, ``Evolutionary Dynamics on
  Degree-Heterogeneous Graphs'', Phys.\ Rev.\ Lett.\ {\bf 96}, 188104 (2006)

\bibitem{SAR08} V. Sood, T. Antal, and S. Redner, ``Voter Models on
  Heterogeneous Networks'', Phys.\ Rev.\ E {\bf 77}, 041121 (2008).

\bibitem{VE08} F. Vazquez and V. M. Eguiluz, ``Analytical Solution of the
  Voter Model on Uncorrelated Networks'', New J. Phys.\ \textbf{10}, 063011
  (2008).

\bibitem{C85} M. J. A. Condorcet, \emph{essai sur l'application de l'analyse
    \'a la probabilit\'e des d\'ecisions rendues \'a la pluralit\'e des voix}
  (L'imprimerie Royale, Paris, France, 1985) [fascimile edition New York:
  Chelsea, 1972].
 
\bibitem{GOF83} B. Grofman, G. Owen, and S. L. Feld, ``Thirteen Theorems In
  Search of the Truth'', Theory and Decision \textbf{15}, 261 (1983).

\bibitem{BR05} R. Boyd and P. J. Richerson, \emph{The Origin and Evolution of
    Cultures}, (Oxford University Press, Oxford, UK, 2005).

\bibitem{CL09} L. Conradt and C. List, ``Group Decision Making in Humans and
  Animals'', Philos.\ Trans.\ Roy.\ Soc.\ London B Biol.\ Sci.\ \textbf{364}
  719 (2009).

\bibitem{SKR2001a} V. Spirin, P. L. Krapivsky, and S. Redner, Phys.\ Rev.\ E
  \textbf{63}, 036118 (2001).

\bibitem{SKR2001b} V. Spirin, P. L. Krapivsky, and S. Redner, Phys.\ Rev.\ E
  \textbf{65}, 016119 (2001).
  
\bibitem{G99} S. Galam, ``Application of Statistical Physics to Politics'',
  Physica A \textbf{274}, 132 (1999).
  
\bibitem{SS00} K. Sznajd-Weron and J. Sznajd, ``Opinion Evolution in Closed
  Community'', Int.\ J. Mod.\ Phys.\ C \textbf{11}, 1157 (2000). 

\bibitem{G02} S. Galam, ``Minority Opinion Spreading in Random Geometry'',
  Eur.\ Phys.\ J. B \textbf{25}, 403 (2002).

\bibitem{D02} D. Stauffer, ``Monte Carlo Simulations of Sznajd Models'',
  J. Artif.\ Soc.\ Soc.\ Simul.\ \textbf{5}, 1 (2002). 

\bibitem{KR03} P. L. Krapivsky and S. Redner, ``Dynamics of Majority Rule in
  Two-State Interacting Spin Systems'', Phys.\ Rev.\ Lett.\ \textbf{90},
  238701 (2003).
  
\bibitem{BO78} C. M. Bender and S. A. Orszag, {\it Advanced Mathematical
    Methods for Scientists and Engineers} (McGraw-Hill, New York, 1978).

\bibitem{CR05} P. Chen and S. Redner, ``Majority Rule Dynamics in Finite
  Dimensions'', Phys.\ Rev.\ E 71, \textbf{036101} (2005).
  
\bibitem{LR07} R. Lambiotte and S. Redner, ``Dynamics of Vacillating
  Voters'', J. Stat.\ Mech.\ L10001, (2007).

\bibitem{LR08} R. Lambiotte and S. Redner, ``Dynamics of Non-Conservative
  Voters''< Europhys.\ Lett.\ \textbf{82}, 18007 (2008).
  
\bibitem{SSP08} F. Slanina, K. Sznajd-Weron, and P. Przyby\l a, ``Some New
  Results on One-Dimensional Outflow Dynamics'', Europhys.\ Lett.
  \textbf{82}, 18006 (2008).

\bibitem{LTH07} R. Lambiotte, S. Thurner, and R. Hanel, ``Unanimity Rule on
  Networks'', Phys.\ Rev.\ E, \textbf{76}, 046101 (2007).
  
\bibitem{G63} R.~J.~Glauber, ``Time‐Dependent Statistics of the Ising
  Model'', J. Math.\ Phys.\ {\bf 4}, 294 (1963).

\bibitem{MR03} M. Mobilia and S. Redner, ``Majority Versus Minority Dynamics:
  Phase Transition in an Interacting Two-State Spin System'', Phys.\ Rev.\ E,
  \textbf{68}, 046106 (2003).

\bibitem{CW12} N. Claidi\`ere and A. Whiten, ``Integrating The Study of
  Conformity and Culture in Humans and Nonhuman Animals.'' Psychol.\ Bull.\
  \textbf{138}, 126 (2012).
  
\bibitem{ML12} T. J. H. Morgan and K. N. Laland, ``The Biological Bases of
  Conformity'', Front.\ Neurosci.\ \textbf{6}, 87 (2012).
  
\bibitem{bA97} D. ben-Avraham, in \emph{Non-equilibrium Statistical Mechanics
    in One Dimension}, ed.\ V. Privman (Cambridge University Press,
  Cambridge, UK, 1997), Chap.\ 2.
  
\bibitem{VR04} F. Vazquez and S. Redner, ``Ultimate Fate of Constrained
  Voters'', J. Phys.\ A {\bf 37}, 8479 (2004).

\bibitem{A77} R. Axelrod, ``The Dissemination of Culture: A Model with Local
  Convergence and Global Polarization'', J. Conflict Res.\ {\bf 41}, 203
  (1977).

\bibitem{AAE96} R. Axtell, R.~Axelrod, J. Epstein, and M. D. Cohen,
  ``Aligning Simulation Models: A Case Study and Results'', Comput.\ Math.\
  Organiz.\ Theory {\bf 1}, 123 (1996).
 
\bibitem{A97} R. Axelrod, \emph{The Complexity of Cooperation} (Princeton
  University Press, Princeton, NJ, 1997).

\bibitem{CMV00} C. Castellano, M. Marsili, and A. Vespignani,
  ``Nonequilibrium Phase Transition in a Model for Social Influence'', Phys.\
  Rev.\ Lett.\ 85, 3536 (2000).

\bibitem{VVC02} D. Vilone, A. Vespignani, and C. Castellano, ``Ordering Phase
  Transition in the One-Dimensional Axelrod Model'', Eur.\ Phys.\ J. B
  \textbf{30}, 399 (2002).
  
\bibitem{KETM03a} K. Klemm K, V. M. Eguiluz, R. Toral, and M. San Miguel,
  ``Nonequilibrium Transitions in Complex Networks: A Model of Social
  Interaction'', Phys.\ Rev.\ E, \textbf{67}, 026120 (2003).

\bibitem{KETM03b} K. Klemm K, V. M. Eguiluz, R. Toral, and M. San Miguel,
  ``Global Culture: A Noise-Induced Transition in Finite Systems'', Phys.\
  Rev.\ E, \textbf{67}, 045101(R) (2003).

\bibitem{VR07} F. Vazquez and S. Redner, ``Non-Monotonicity and Divergent
  Time Scale in Axelrod Model Dynamics'', Europhys.\ Lett. \textbf{78}, 18002
  (2007).

\bibitem{L63} E. N. Lorenz, ``Deterministic Nonperiodic Flow'', J. Atmos.\
  Sci.\ \textbf {20}, 130 (1963).

\bibitem{ML75} R. M. May and W. J. Leonard, ``Nonlinear Aspects of
  Competition Between Three Species'', SIAM J. Appl.\ Math.\ \textbf{29}, 243
  (1975).

\bibitem{PN99} See e.g., A. S. Perelson and P. W. Nelson, ``Mathematical
  Analysis of HIV-1 Dynamics in Vivo'', SIAM Review \textbf{41}, 3 (1999).

\bibitem{WDAN02} G. Weisbuch, G. Deffuant, F. Amblard, and J. P. Nadal,
  ``Meet, Discuss, and Segregate!'', Complexity {\bf 7}, 55 (2002).

\bibitem{HK02} R. Hegselmann, U. Krause, ``Opinion Dynamics and Bounded
  Confidence Models, Analysis, and Simulation'', J. Artif.\ Soc.\ Soc.\
  Simul.\ \textbf{5}, 3 (2002).

\bibitem{BKR03} E. Ben-Naim, P.~L.~Krapivsky, and S. Redner, ``Bifurcations
  and Patterns in Compromise Processes'', Physica D {\bf 183}, 190 (2003).

  
\end{thebibliography}
\end{document}